%% file: S2COSMOS_accepted.tex
\documentclass[11pt,a4paper]{emulateapj}
\usepackage{graphicx}
\usepackage{textcomp}
\usepackage{amsmath}
\usepackage{amssymb}
\usepackage{gensymb}
\usepackage{rotating}
\usepackage{epsfig}
\usepackage{amsmath}
\usepackage{longtable}
\usepackage{color}
 \def\sigsubm{\sigma_{850\mu\mathrm{m}}} 
  
 \def\fluxsubmint{\mathrm{S}_{850\mu\mathrm{m}}^{\mathrm{Intr.}}} 
 \def\mjpb{mJy\,beam$^{-1}$}
 \def\gs{\mathrel{\raise0.35ex\hbox{$\scriptstyle >$}\kern-0.6em\lower0.40ex\hbox{{$\scriptstyle \sim$}}}}
 \def\ls{\mathrel{\raise0.35ex\hbox{$\scriptstyle <$}\kern-0.6em\lower0.40ex\hbox{{$\scriptstyle \sim$}}}}

 \def\Msol{\mathrel{\rm M_{\odot}}}
 \def\Lsol{\mathrel{\rm L_{\odot}}}
 \def\Msolyr{\mathrel{\rm M_{\odot}\,yr^{-1}}}
 
 \def\Wm2{\,\hbox{W}\,\hbox{m}^{-2}}
 \def\gsim{\mathrel{\raise0.35ex\hbox{$\scriptstyle >$}\kern-0.6em\lower0.40ex\hbox{{$\scriptstyle \sim$}}}}
 \def\lsim{\mathrel{\raise0.35ex\hbox{$\scriptstyle <$}\kern-0.6em\lower0.40ex\hbox{{$\scriptstyle \sim$}}}}
 
 \def\pc{\%}

\lefthead{Simpson et al.}  \righthead{S2COSMOS: An EAO\,/\,JCMT survey of 1147 sub-millimeter sources}

\begin{document}

\title{The East Asian Observatory SCUBA--2 survey of the COSMOS field: unveiling 1147 bright sub--millimeter sources across 2.6\,square degrees}

\author{
J.\,M.\ Simpson,\altaffilmark{1}
Ian Smail,\altaffilmark{2}
A.\,M.\ Swinbank,\altaffilmark{2}
S.\,C.\ Chapman,\altaffilmark{3}
Chian-Chou\ Chen,\altaffilmark{4}
J.\,E.\ Geach,\altaffilmark{6}
Y.\ Matsuda,\altaffilmark{7,8}
R.\ Wang,\altaffilmark{9}
Wei-Hao Wang,\altaffilmark{1}
Y.\ Yang,\altaffilmark{10}
Y.\ Ao,\altaffilmark{11}
R.\ Asquith,\altaffilmark{12}
N.\,Bourne,\altaffilmark{5}
R.\ T.\ Coogan,\altaffilmark{13}
K.\,Coppin,\altaffilmark{6}
B.\, Gullberg,\altaffilmark{2}
N.\,K.\ Hine,\altaffilmark{6}
L.\,C.\ Ho,\altaffilmark{9,14}
H.\,S.\ Hwang,\altaffilmark{10}
R.\,J.\ Ivison,\altaffilmark{4,5}
Y.\ Kato,\altaffilmark{7}
K.\ Lacaille,\altaffilmark{15}
A.\,J.\,R.\ Lewis,\altaffilmark{5}
D.\ Liu,\altaffilmark{16}
M.\,J.\ Micha{\l}owski,\altaffilmark{17}
I.\ Oteo,\altaffilmark{4,5}
M.\ Sawicki,\altaffilmark{18}
J.\ Scholtz,\altaffilmark{2}
D.\ Smith,\altaffilmark{6}
A.\,P.\,Thomson,\altaffilmark{19}
J.\ L.\ Wardlow,\altaffilmark{20}}

\setcounter{footnote}{0}
\altaffiltext{1}{EACOA fellow: Academia Sinica Institute of Astronomy and Astrophysics, No.\ 1, Sec.\ 4, Roosevelt Rd., Taipei 10617, Taiwan; email: jsimpson@asiaa.sinica.edu.tw}
\altaffiltext{2}{Centre for Extragalactic Astronomy, Department of Physics, Durham University, South Road, Durham DH1 3LE, UK} 
\altaffiltext{3}{Department of Physics and Atmospheric Science, Dalhousie University, Halifax, NS B3H 3J5 Canada}
\altaffiltext{4}{European Southern Observatory, Karl Schwarzschild Strasse 2, Garching, Germany}
\altaffiltext{5}{Institute for Astronomy, University of Edinburgh, Royal Observatory, Blackford HIll, Edinburgh EH9 3HJ, UK}
\altaffiltext{6}{Centre for Astrophysics Research, Science and Technology Research Institute, University of Hertfordshire, Hatfield AL10 9AB, UK}
\altaffiltext{7}{National Astronomical Observatory of Japan, 2-21-1 Osawa, Mitaka, Tokyo, 181-8588, Japan}
\altaffiltext{8}{The Graduate University for Advanced Studies (SOKENDAI), Osawa, Mitaka, Tokyo, 181-8588, Japan}
\altaffiltext{9}{Kavli Institute for Astronomy and Astrophysics, Peking University, Beijing 100871, China}
\altaffiltext{10}{Korea Astronomy and Space Science Institute, 776 Daedeokdae-ro, Yuseong-gu, Daejeon 34055, Republic of Korea}
\altaffiltext{11}{Purple Mountain Observatory \& Key Laboratory for Radio Astronomy, Chinese Academy of Sciences, 8 Yuanhua Road, Nanjing 210034, China}
\altaffiltext{12}{School of Physics \& Astronomy, University of Nottingham, Nottingham NG7 2RD, UK}
\altaffiltext{13}{Astronomy Centre, Department of Physics and Astronomy, University of Sussex, Brighton BN1 9QH, UK}
\altaffiltext{14}{Department of Astronomy, Peking University, Beijing, 100087, China}
\altaffiltext{15}{Department of Physics and Astronomy, McMaster University, Hamilton, ON L8S 4M1 Canada}
\altaffiltext{16}{Max Planck Institute for Astronomy, K\"{o}nigstuhl 17, D-69117 Heidelberg, Germany}
\altaffiltext{17}{Astronomical Observatory Institute, Faculty of Physics, Adam Mickiewicz University, ul.~S\l{}oneczna 36, 60-286 Pozna\'n, Poland}
\altaffiltext{18}{Department of Astronomy and Physics, Saint Mary’s University, Halifax, NS B3H 3C3, Canada}
\altaffiltext{19}{The University of Manchester, Oxford Road, Manchester, M13 9PL, UK}
\altaffiltext{20}{Physics Department, Lancaster University, Lancaster, LA1 4YB, UK}

\begin{abstract}
We present sensitive 850\,$\mu$m imaging of the COSMOS field using 640\,hr of new and archival observations taken with SCUBA--2 at the East Asian Observatory's James Clerk Maxwell Telescope. The SCUBA--2 COSMOS survey (S2COSMOS) achieves a median noise level of $\sigma_{850\mu{\mathrm{m}}}$\,=\,1.2\,mJy\,beam$^{-1}$ over an area of 1.6 sq.~degree ({\sc {main}}; {\it{Hubble Space Telescope\,/\,Advanced Camera for Surveys}} footprint), and $\sigma_{850\mu{\mathrm{m}}}$\,=\,1.7\,mJy\,beam$^{-1}$ over an additional 1 sq.\ degree of supplementary ({\sc supp}) coverage. We present a catalogue of 1020 and 127 sources detected at a significance level of $>$\,4\,$\sigma$ and $>$\,4.3\,$\sigma$ in the {\sc main} and {\sc supp} regions, respectively, corresponding to a uniform 2\,$\pc$ false--detection rate. We construct the single--dish 850\,$\mu$m number counts at $S_{850}$\,$>$\,2\,mJy and show that these S2COSMOS counts are in agreement with previous single-dish surveys, demonstrating that degree--scale fields are sufficient to overcome the effects of cosmic variance in the $S_{850}$\,$=$\,2--10\,mJy population. To investigate the properties of the galaxies identified by S2COSMOS sources we measure the surface density of near-infrared--selected galaxies around their positions and identify an average excess of 2.0\,$\pm$\,0.2 galaxies within a 13$''$ radius ($\sim$\,100\,kpc at $z$\,$\sim$\,2). The bulk of these galaxies represent near--infrared-selected SMGs and\,/\,or spatially--correlated sources and lie at a median photometric redshift of $z$\,=\,2.0\,$\pm$\,0.1. Finally, we perform a stacking analysis at sub--millimeter and far--infrared wavelengths of stellar--mass-selected galaxies ($M_{\star}$\,=\,10$^{10}$--10$^{12}$\,$\Msol$) from $z$\,=\,0--4, obtaining high-significance detections at 850\,$\mu$m in all subsets (signal--to--noise ratio, SNR\,=\,4--30), and investigate the relation between far--infrared luminosity, stellar mass, and the peak wavelength of the dust SED. The publication of this survey adds a new deep, uniform sub--millimeter layer to the wavelength coverage of this well--studied COSMOS field.
\end{abstract}


\section{Introduction}
\label{sec:intro}
Understanding the evolution of galaxies over cosmic time and, thus, the growth of stellar mass in the Universe, is a fundamental objective of modern astrophysics. The importance of observations at far--infrared wavelengths for the study of galaxy evolution has been clear since the discovery that the integrated emission from all galaxies in the Universe, the extragalactic background, has a comparable intensity at optical and infrared wavelengths \citep{Puget96, Fixsen98, Hauser98} -- i.e.\ approximately half of the total energy that is radiated by galaxies in the ultraviolet\,/\,optical is reprocessed by dust and emitted in the far--infrared. The {\it Infrared Astronomical Satellite} ({\it IRAS}; \citealt{Neugebauer84}) all-sky survey provided the first census of obscured activity, demonstrating that local galaxies emit, on average, one third of their bolometric luminosity at infrared wavelengths \citep{Soifer91}. {\it IRAS} also established the presence of a population of galaxies whose bolometric luminosity is dominated by their emission at far--infrared wavelengths (for a review see \citealt{Sanders96}). The most luminous of these galaxies are termed Ultra Luminous Infrared Galaxies (ULIRGs) and have total far--infrared luminosities $>$\,10$^{12}$\,$\Lsol$ that arise, primarily, from the reprocessing of ultraviolet emission associated with intense star formation by dust in the interstellar medium (e.g.\ \citealt{Lutz98}). Despite hosting regions of strong star formation ($\gsim$\,100\,$\Msolyr$) the low volume density of ULIRGs means that they represent a negligible component ($\ll$\,1\,$\pc$) of the integrated bolometric luminosity of galaxies at low redshift. 

It is now two decades since the first extragalactic surveys at sub--millimeter (sub--mm) wavelengths isolated a cosmologically--significant population of sub--millimeter sources at high redshift \citep{Smail97,Hughes98,Barger98,Lilly99}. These 850\,$\mu$m surveys, undertaken with Sub--mm Common User Bolometer Array (SCUBA) on the 15-m James Clerk Maxwell Telescope (JCMT), uncovered the bright--end ($S_{850}$\,=\,5--15\,mJy) of the sub--millimeter galaxy (SMG; $S_{850}$\,$>$\,1\,mJy) population and demonstrated that the space density of systems with ULIRG--like luminosities increases by three orders of magnitude towards high redshift (e.g.\ \citealt{Smail97}). Subsequent efforts to obtain sensitive sub--mm imaging over wider areas typically uncovered samples of $\sim$\,100 sources (e.g.\ \citealt{Scott02, Coppin06, Weiss09}) that, when twinned with multi-wavelength follow--up campaigns \citep{Biggs11}, confirmed that SMGs lie at a typical redshift of $z$\,$\sim$\,2.5 (\citealt{Chapman05,Simpson14}); have star--formation rates of $\gsim$\,300\,$\Msolyr$ (\citealt{Magnelli12,Swinbank13}); contain vast reservoirs of molecular gas ($M_{\mathrm{gas}}$\,$\sim$\,10$^{10}$\,$\Msol$; \citealt{Bothwell13}); often host an Active Galactic Nucleus \citep{Alexander05a,Pope08,Wang13}; and, crucially, contribute $\sim$\,20--30\,$\pc$ to the cosmic star-formation rate density over a wide range in lookback time (e.g.\ \citealt{Swinbank13,Cowie17}). Thus, while infrared--dominated systems are negligible sources of star formation in the local Universe they represent a crucial component of the galaxy population at higher redshift.    

Despite initial efforts to characterize SMGs, the relatively small number of known sources meant that key properties regarding their connection to other galaxy populations (e.g.\ environment, clustering) remained poorly constrained (\citealt{Hickox12}). The launch of the {\it Herschel} satellite \citep{Pilbratt10} and the subsequent wide-field surveys with the PACS and SPIRE instruments (operating at 70--500\,$\mu$m) drastically increased the number of known far--infrared--luminous systems at cosmologically significant redshifts (e.g.\ \citealt{Roseboom12,Oliver12,Magnelli13,Bourne16}; but see also \citealt{Vieira10}). In particular, a suite of extragalactic surveys mapped $\sim$\,1000\,sq.\ degree to varying sensitivity, primarily at 250--500\,$\mu$m, and enabled the identification and characterization of infrared emission out to moderate redshift ($z$\,$\sim$\,1--2; e.g.\ \citealt{Dunne11,Gruppioni13,Eales18}). However, at high redshift robust detections are typically limited to hyper--luminous (e.g.\ \citealt{Asboth16,Ivison16}) or gravitationally--lensed sources (e.g.\ \citealt{Negrello10}), due to a combination of both the coarse resolution of long--wavelength {\it Herschel} imaging ($\sim$\,25--35$''$ FWHM) and a rapidly evolving $k$--correction with redshift.

The $k$--correction is defined as the change in apparent luminosity of a source in a fixed waveband due to the effect of redshift. For a source that is observed in the Rayleigh--Jeans regime an increase in redshift shifts the peak of the dust spectral energy distribution (SED) through the waveband, resulting in an initial ``brightening'' that counters the effect of cosmological dimming. This ``negative'' $k$--correction is strong at sub--mm wavelengths and, under the assumption of a constant dust temperature, means that observations conducted at $\sim$\,850\,$\mu$m provide an almost distance--independent selection of infrared sources from $z$\,=\,0\,--\,7 (e.g.\ \citealt{Blain02}). For this reason, flux limited observations conducted in the classical sub--mm\,/\,mm regime remain the most effective way to systematically study the infrared--bright galaxy population at high redshift. 

The SCUBA--2 Cosmology Legacy Survey (S2CLS; \citealt{Geach17}) represents the largest area, sensitive survey of the sub--mm sky that has been undertaken to date. The 850\,$\mu$m component of the survey is comprised of 4 sq.\ degree of sensitive imaging, distributed over seven extragalactic survey fields, and was obtained with the currently unparalleled SCUBA--2 \citep{Holland13} camera at the JCMT. Key targets for S2CLS included the UKIDSS Ultra Deep Survey (UDS) and the Cosmological Evolution Survey (COSMOS) fields, representing the two premier degree--scale extragalactic survey regions. The planned S2CLS observations of the UDS were completed, yielding a large sample of sub-mm sources across 0.9 sq.\ degree for further study (e.g.\ \citealt{Smail14,Simpson15b,Chen16b,Wilkinson17,Stach18}) and have given tentative insights into the evolutionary connection between sub--mm sources and other galaxy populations \citep{Wilkinson17}. However, the COSMOS component of S2CLS was not fully completed and this resulted in an inhomogeneous map at 850\,$\mu$m, with particularly shallow coverage across one half of the field (see \citealt{Geach17}). The COSMOS field has the richest set of ancillary data of any degree--scale field, with a cornucopia of imaging at nm-to--cm wavelengths, and has been the target of a number of extensive spectroscopic surveys (e.g.\ \citealt{Lilly07,Hasinger18}). To compliment the existing data in this field and connect obscured activity at high redshift with the well--studied unobscured galaxy population requires a complete survey of the whole field at 850\,$\mu$m.

Here, we present the completed, homogeneous survey of the COSMOS field with SCUBA--2 undertaken as part of the East Asian Observatories (EAO) Large Program series. The SCUBA--2 COSMOS survey (S2COSMOS) aims to provide a deep, contiguous image of the full COSMOS field at 850\,$\mu$m, by adding 223\,hr of observations to the 416\,hr of archival coverage that was primarily obtained as part of S2CLS (\citealt{Geach17}; see also \citealt{Casey13}). In principle COSMOS represents a 2 sq.\ degree region of sky, but significant variations exist between the footprints of different multiwavelength datasets. In this paper we define an S2COSMOS {\sc main} survey area that corresponds to the 1.6\,sq.\,degree region of the field that was imaged (\citealt{Koekemoer07}) with the Advanced Camera for Surveys (ACS) onboard the {\it Hubble Space Telescope} ({\it HST}). This {\sc main} region broadly represents the intersection between deep surveys of the field at optical--to--near-infrared wavelengths (e.g.\ \citealt{Koekemoer07,Sanders07,McCracken12,Laigle16}), and thus the region of the map with high-quality photometric redshift estimates that are key to further study of the SMG population. As sensitive sub--mm imaging does exist beyond this {\sc main} region we also define a S2COSMOS {\sc supplementary} ({\sc supp}) region that is contiguous to, but extends beyond, the central {\sc main} survey.

In this paper we present the observations, data reduction and analysis of the S2COSMOS survey, and release a catalogue of extracted sources at 850\,$\mu$m. The paper is structured as follows. In \S\,2 we present our survey strategy, observations and data reduction. In \S\,3 we describe our source extraction procedure, along with statistical tests to determine the fidelity of the resulting catalog. In \S\,4 we discuss the properties of the SCUBA--2 detections and present number counts for the 850--$\mu$m--luminous population. Furthermore, we combine our deep 850\,$\mu$m imaging with multiwavelength imaging of the COSMOS field to study the average properties (e.g.\ star-formation rate, dust temperature, gas mass) of mass--selected sources from $z$\,=\,0--4. Our conclusions are given in \S\,5. Throughout this work we adopt the AB magnitude system, a \citet{Chabrier03} stellar initial mass function (IMF) and a cosmology with with $H_{\rm 0}$\,=\,67.8\,km\,s$^{-1}$\,Mpc$^{-1}$, $\Omega_{\Lambda}$\,=\,0.69, and $\Omega_{\rm m}$\,=\,0.31 \citep{Planck14}.


\begin{figure*}
  \centering
  \includegraphics[width=0.65\textwidth]{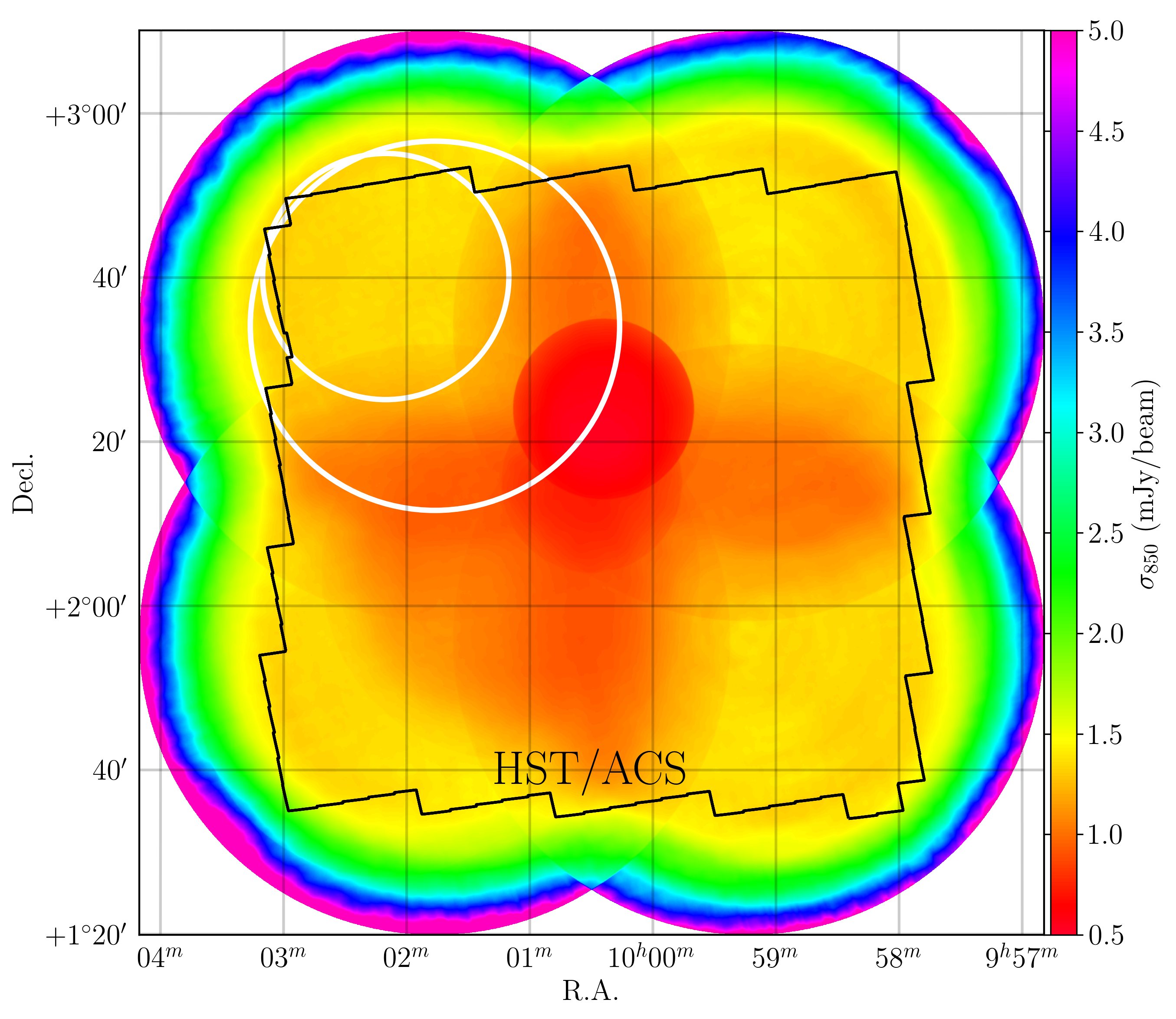} 
  \includegraphics[width=0.6\textwidth]{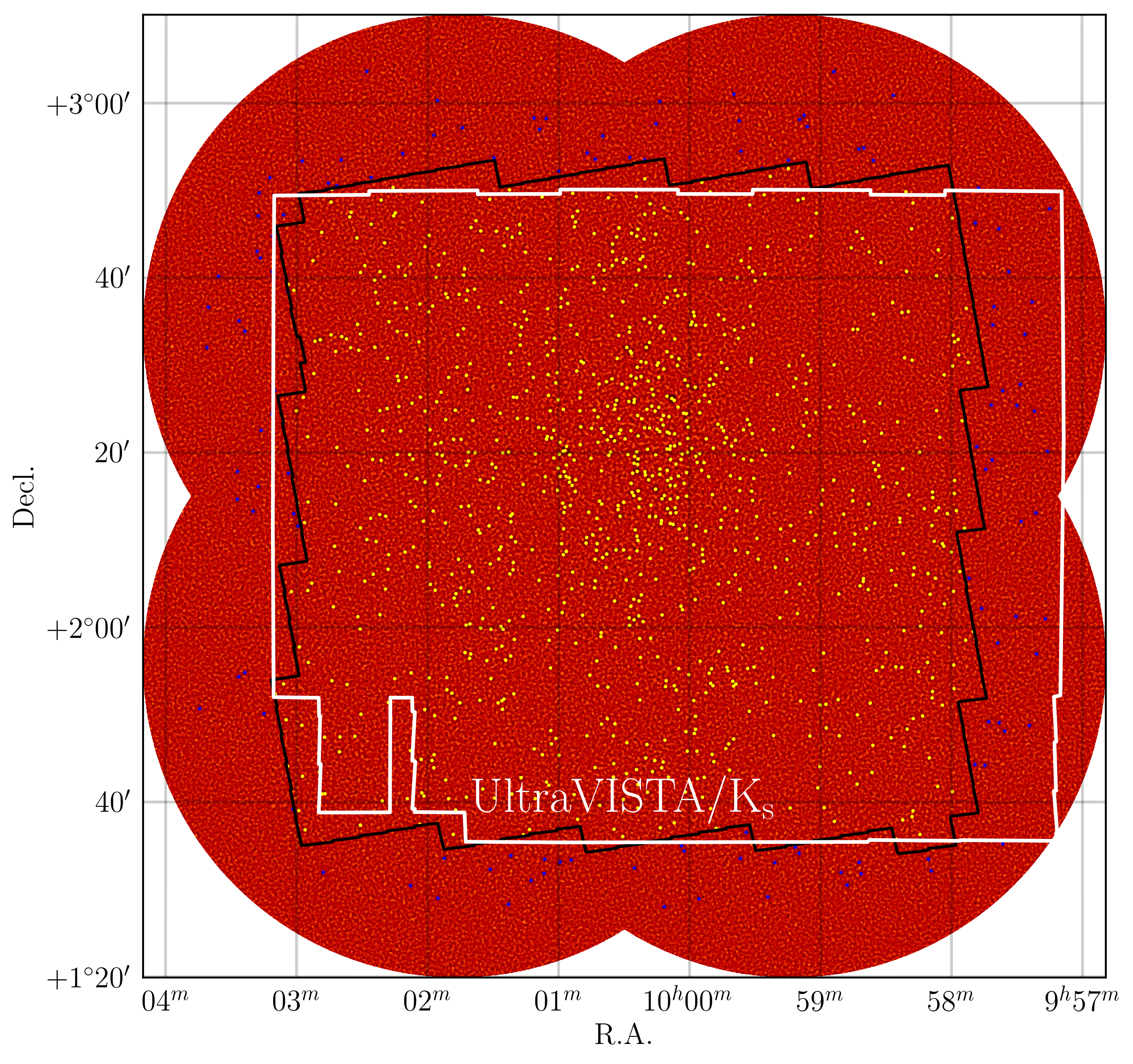} 
  \caption{ Coverage maps of the COSMOS field from S2COSMOS, our legacy survey with SCUBA--2 at the JCMT. {\textit{Top:}} The instrumental noise map of the COSMOS field at 850\,$\mu$m achieved by combining 223\,hr of S2COSMOS observations, undertaken as an EAO large program, with archival imaging of the field (\citealt{Geach17}, \citealt{Casey13}). White circles represent examples of the S2COSMOS observing strategy and show the location and nominal area coverage of the PONG\,2700 and PONG\,1800 scan patterns; these scans were repeated in each quadrant of the field at equidistant positions from the field centre. The S2COSMOS map reaches a median 1--$\sigma$ sensitivity of 1.3\,mJy\,beam$^{-1}$ over the nominal 2\,sq.\,degree COSMOS field. The black outline represents the {\textit{HST}}\,/\,ACS footprint in COSMOS and defines the S2COSMOS {\sc main} survey region (1.6 sq.\,degree; median $\sigsubm$\,=\,1.2\,mJy\,beam$^{-1}$). {\textit{Bottom:}} The signal-to-noise ratio map at 850\,$\mu$m from our S2COSMOS survey. We identify 1020 sources (yellow circles) at a detection significance of $>$\,4\,$\sigma$ within the {\sc main} survey area. A further 127 sources (blue circles) are identified at a detection significance of $>$\,4.3\,$\sigma$ in the S2COSMOS supplementary {\sc{supp}} region (median $\sigma_{850}$\,$<$\,3\,mJy\,beam$^{-1}$). Overall, the S2COSMOS survey provides a uniquely large sample of 1147 obscured starbursts with deep, multiwavelength coverage from X-ray to radio wavelengths.
    } 
\label{fig:fieldplan}
\end{figure*}

\section{Observations and Data Reduction}
\subsection{Observations}
Observations for the S2COSMOS project were carried out between Jan.\ 2016 and Jun.\ 2017 using the SCUBA--2 instrument \citep{Holland13} on the JCMT. Although SCUBA--2 observes simultaneously at both 450\,$\mu$m and 850\,$\mu$m we only present here the 850\,$\mu$m data; the 450\,$\mu$m data will be analyzed in future work. Data were obtained in ``good'' weather, corresponding to a median opacity of $\tau_{\mathrm{225GHz}}$\,=\,0.06 (0.04-0.10; 10$^{\mathrm{th}}$--90$^{\mathrm{th}}$ percentile), with conditions monitored via observations of the 183\,GHz water line with the JCMT water vapor radiometer \citep{Dempsey13}. Individual observations were limited to an integration time of $\sim$\,40\,minutes to allow accurate monitoring of conditions and were interspersed with regular pointing observations. Elevation constraints of $>$\,30 and $<$\,70\,degrees were imposed to ensure sufficiently low airmass and to account for the demands of the scan patterns on the telescope tracking. As such, the S2COSMOS data were taken with the field at a median elevation of 56\,degrees (39--68\,degrees; 10$^{\mathrm{th}}$--90$^{\mathrm{th}}$ percentile).

Observations for S2COSMOS were conducted using the SCUBA--2 {\sc PONG--1800} and {\sc PONG--2700} observing strategies (see \citealt{Chapin13}), which provide a uniform coverage over circular regions of radius 15$'$ and 22.5$'$, respectively. To map the full 2 sq.\,degree COSMOS field we adopt the observing strategy used in observations of the field taken as part of S2CLS \citep{Geach17}, the forerunner to our S2COSMOS survey. Principally, data were obtained using four PONG--2700 scans that were located equidistant from the centre of the COSMOS field (see Figure\,1). These PONG--2700 scans provide coverage of the full 2 sq.\ degree COSMOS field but result in inhomogeneous coverage, with higher sensitivity achieved where the scans overlap (see Figure~\ref{fig:fieldplan}). To improve the homogeneity of the final map we also obtained observations in the smaller footprint, PONG--1800 scan pattern, again centered in the four corners of the COSMOS field. Observations were actively managed to ensure that the sensitivity across the field remained close to uniform. Overall 223\,hr of observations were obtained, using the PONG-2700 and PONG--1800 scans in a ratio of five--to--one in terms of total exposure time. 

The COSMOS field has been the target of repeated observations with SCUBA--2 prior to the S2COSMOS survey and we utilize these publicly--available data here. All relevant observations were retrieved from the Canadian Astronomy Data Center (CADC) and processed and analyzed in an identical manner to our bespoke S2COSMOS data. The archival imaging consists of observations undertaken with SCUBA--2 in median $\tau_{\mathrm{wvr}}$\,=\,0.06 (0.04--0.09), median elevation 54\,degrees (38--67\,degrees), and the PONG mapping strategy (radius\,=\,7.5--30$'$). The bulk of the archival data (85\,\%) was obtained as part of S2CLS \citep{Geach17} with the remaining observations conducted in time allocated to the University of Hawaii (see \citealt{Casey13}). 

Overall, we consider 223\,hr of observations with SCUBA--2 that were undertaken as part of S2COSMOS and 416\,hrs of archival imaging to create a 640\,hr legacy wide-field 850\,$\mu$m map of the COSMOS field.

\subsection{Data Reduction}
\label{subsec:reduction}
The SCUBA--2 observations considered here were reduced using the {\it Dynamical Iterative Map Maker} ({\sc dimm}) within the {Sub-Millimeter Common User Facility} ({\sc smurf}), which is provided as part of the {\sc starlink} software suite \citep{Chapin13}. Full details of the data reduction procedure employed by {\sc dimm} are provided in \citet{Chapin13} but we give a brief overview of the process here. 

Each independent, $\sim$\,40\,min observation with SCUBA-2 is reduced separately, with the raw data first undergoing a number of pre-processing steps. During this pre-processing stage the raw data from each of the four SCUBA--2 sub-arrays is concatenated into a single time-stream and down-sampled to a rate that matches the 2$''$ pixel-scale adopted in this work. The data are flat-fielded using fast-flat scans that bracket each individual observation resulting in data in units of pW, and a linear fit to each timestream is used to subtract a baseline level. Any spikes in each time-stream are removed by considering a box-car width of 50 time--slices and a spike threshold of 10\,$\sigma$. Sudden steps in each time-stream are corrected by subtracting the estimated step--height from the affected data and any gaps in the resulting data are filled using linear interpolation of 50 preceding/following time--slices.  

After pre-processing, {\sc dimm} enters an iterative stage where a model comprised of common-mode signal, astronomical signal, and noise is fit to each time-stream. The common-mode signal is calculated independently for each sub-array and the best-fit model is removed from the time-stream. Next, an extinction correction is applied based on the atmospheric opacity as monitored by the JCMT water vapor monitor, and a high-pass filter is adopted to remove data corresponding to spatial scales above 200$''$. The time-stream data are projected onto a pre-defined pixel grid that is kept constant for all observations and the astronomical signal is estimated, inverted back to a time-stream, and then subtracted from all bolometers. The noise of each bolometer is estimated by considering the residuals after subtracting all other signal and is used to estimate the pixel-by-pixel instrumental noise in the final map; the noise estimate includes the contribution from instrument and atmospheric effects and we refer to this as SCUBA-2 instrumental noise throughout. The entire process is repeated for a maximum of 20 iterations and curtailed when the convergence criterion is satisfied ($\Delta$\,$\chi^2$\,$<$\,0.05). 

The data reduction procedure provides a set of individual maps that can be combined to create a mosaic. Before stacking these individual scans we must consider that the maps have different pointing centers and nominal radii. In particular, while each reduced map achieves a uniform noise level over a nominal radius the true coverage extends over a significantly wider area, albeit at rapidly decreasing sensitivity and fidelity due to the limited number of bolometers that target this region. To investigate the reliability of the extended, shallower coverage, we empirically measure the noise for each scan pattern in radially-averaged annuli from the centre of each map and compare this to the expected instrumental noise. The measured noise profile is found to be in good agreement with the expected instrument noise at $<$\,1.5\,$\times$ the nominal map radius of the recipe, and as such each individual map is cropped at this threshold. The individual maps are combined on a pixel-by-pixel basis using inverse--variance weighting and rejecting any outliers that lie at $\pm$\,6\,$\sigma$ from the median. Note that during the data reduction stage the bolometer time-streams from each observation were projected onto a consistent reference frame and, as such, no further re-projection or astrometric correction was required to combine the individual maps into a single mosaic.

Finally, we apply three additional post-processing steps to the 850\,$\mu$m mosaic. First, to improve sensitivity to point-source emission we apply a matched--filter to the map using {\sc starlink}\,/\,{\sc picard} and the recipe {\it scuba2{\textunderscore}matched{\textunderscore}filter}. The matched-filtering consists of two steps: large-scale residual noise is first removed by smoothing the image with a Gaussian of FWHM\,=\,30$''$ and subtracting the result from the original image; then the image is convolved with the PSF of the telescope (\citealt{Dempsey13}; corrected for the prior smoothing step) to provide optimal sensitivity to point source emission. Secondly, we adopt the standard SCUBA--2\,850\,$\mu$m flux conversion factor (FCF) of $537$\,Jy\,beam$^{-1}$\,pW$^{-1}$ to convert the map into units of flux density. This FCF value was derived by considering historical data for over 500 observations of calibrators (see \citealt{Dempsey13}) and the absolute calibration uncertainty is expected to be $<$\,8$\pc$. Finally, we account for the loss of flux density introduced during the filtering steps employed in the data reduction. To measure the flux loss due to filtering we inject 1000 simulated point sources into the timestream data with flux densities of 0.5--20\,Jy. We determine that an upwards correction of 13\,$\pc$ is required to correct for flux loss due to filtering effects and we apply this to the S2COSMOS maps (see also \citealt{Geach17}).

\begin{figure}
  \centering
  \includegraphics[width=0.48\textwidth]{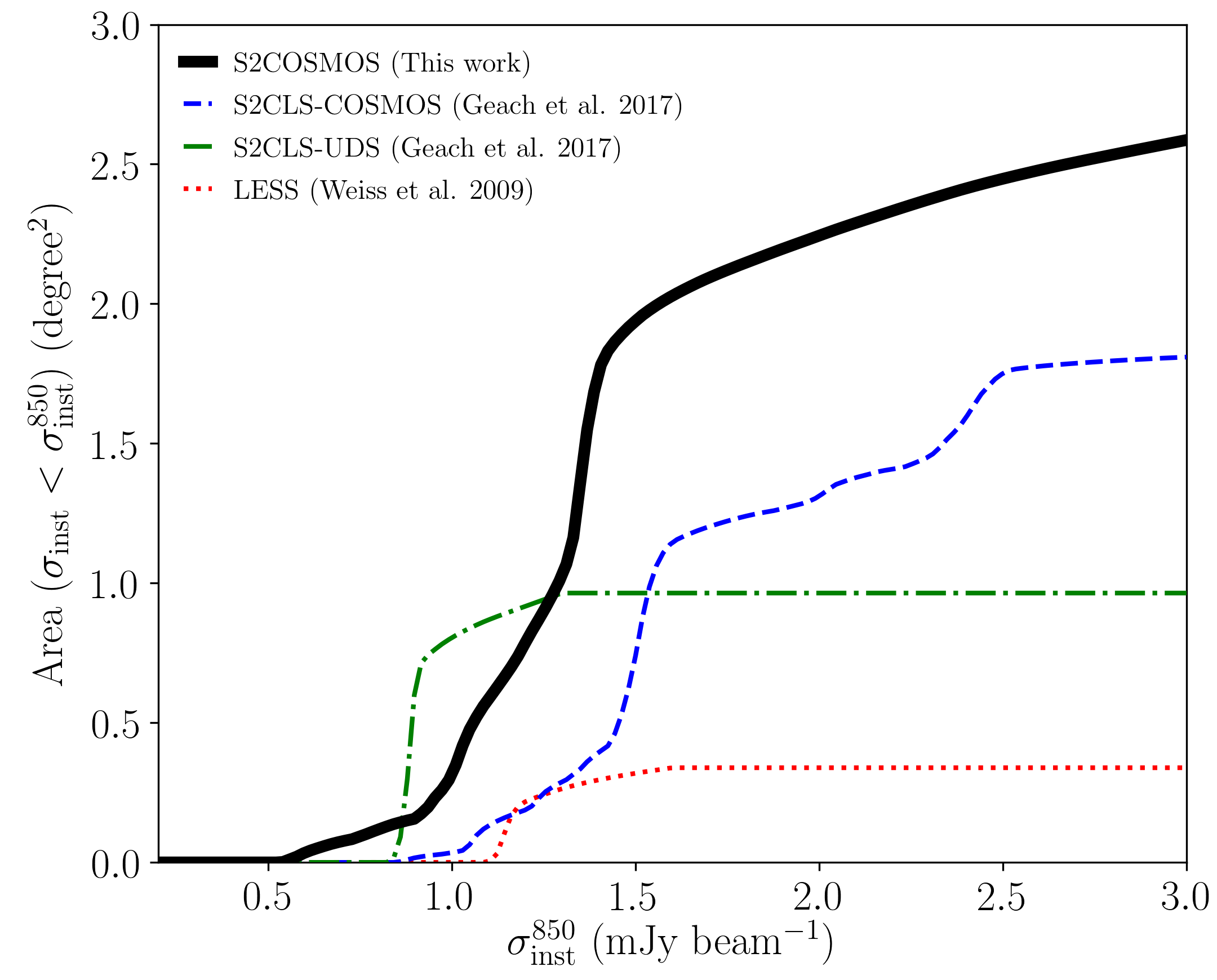} 
  \caption{ The cumulative area of the S2COSMOS survey as a function of instrumental sensitivity, compared to previous surveys with SCUBA--2\,/\,850\,$\mu$m (S2CLS; \citealt{Geach17}) and LABOCA\,/\,870\,$\mu$m (LESS; \citealt{Weiss09}). S2COSMOS builds upon S2CLS--COSMOS (dashed line; \citealt{Geach17}) to achieve a median sensitivity of $\sigsubm$\,=\,1.2\,\mjpb\ over the {\it HST}\,/\,ACS COSMOS region, dramatically improving the depth and homogeneity of the 850\,$\mu$m imaging in this key extragalactic survey field.
  } 
\label{fig:rmsstruct}
\end{figure}

\subsection{Properties of the S2COSMOS map}
\subsubsection{Coverage Map}
In Figure~\ref{fig:fieldplan} we show the S2COSMOS coverage map of the COSMOS field represented in terms of the achieved instrumental sensitivity and the point-source signal--to--noise ratio. As described in \S~\ref{subsec:reduction}, coverage of the field is achieved by mosaicking circular maps with varying radii and pointing centers. As a result the instrumental sensitivity varies across the final map and we discuss this inhomogeneity here. The instrumental noise is typically lower in regions where the scan patterns overlap, with the deepest regions of the map reaching $\sigma_{850}$\,=\,0.53\,mJy\,beam$^{-1}$, close to the expected confusion noise (see \S\,\ref{subsec:confusion}, but also \citealt{Blain02}). The noise increases rapidly in the outskirts of the map, where coverage is limited to regions of telescope over-scan and the resulting integration time per pixel is lower. The instrumental noise in these outer regions increases to $\sigma_{850}$\,$\lsim$\,5\,mJy, although we note that we do not consider the lowest sensitivity regions for source extraction.

The survey area of the S2COSMOS map as a function of the instrumental noise is shown in Figure~\ref{fig:rmsstruct}. For comparison, we show the noise profile of the 850\,$\mu$m imaging of the UDS and COSMOS fields taken from S2CLS \citep{Geach17}, and the LABOCA Survey of the Extended {\it Chandra} Deep Field (LESS; \citealt{Weiss09}). S2COSMOS builds upon the S2CLS--COSMOS survey to achieve an instrumental sensitivity of $\sigma^{\mathrm{inst}}_{850}$\,=\,0.5--2.4\,mJy\,beam$^{-1}$ over 2.2 sq.\,degree, a significant improvement upon the observations taken as part of S2CLS; S2CLS--COSMOS mapped 2.2\,sq.\,degree to a depth of $\sigma^{\mathrm{inst}}_{850}$\,=\,0.8--4.5\,mJy\,beam$^{-1}$, with subsequent source extraction limited to a 1.3 sq.\,degree region ($\sigma^{\mathrm{inst}}_{850}$\,$<$2\,\mjpb). Furthermore, S2COSMOS provides an improved uniformity in the noise level across the central regions of the COSMOS field, relative to S2CLS--COSMOS, as demonstrated by the sharp rise in the total area surveyed at $\sigma^{\mathrm{inst}}_{850}$\,$\lsim$\,1.4\,\mjpb (see Figure~\ref{fig:rmsstruct}). 

The S2COSMOS {\sc main} survey represents a 1.6\,sq.\,degree region of the S2COSMOS map with a median 1--$\sigma$ instrumental sensitivity of 1.2\,mJy\,beam$^{-1}$ (16-84th percentile: 1.0--1.4\,mJy\,beam). The {\sc supp} region provides a further 1\,sq.\,degree of 850\,$\mu$m imaging, at a median 1--$\sigma$ instrumental sensitivity of 1.7\,mJy\,beam$^{-1}$ (16-84th percentile: 1.4--2.5\,mJy\,beam). An upper limit of $<$\,3\,mJy\,beam$^{-1}$ for the {\sc supp} regions was chosen to increase the total S2COSMOS survey area for the rarest, most-luminous sources (S$_{850}$\,$\gsim$\,10\,mJy; see \citealt{Geach17}), while balancing the effect of flux boosting and an increasing false-detection rate in these lower sensitivity regions (see \S\,\ref{subsec:sourceext}). For reference, the {\sc main} and {\sc supp} survey areas correspond to a survey volume of 9.7 and 5.6\,$\times$\,10$^{7}$\,Mpc$^3$, respectively, assuming a typical redshift range of $z$\,=\,0.5--6.0 for the sub-millimeter--luminous population (e.g.\ \citealt{Simpson14,Strandet16}). Imaging at near--\,/\,mid--infrared wavelengths is imperative for understanding the physical properties of 850\,$\mu$m--selected sources (e.g.\ \citealt{Simpson17}) and we note that 98\,$\pc$ of the {\sc supp} sources fall within the {\it Spitzer}\,/\,IRAC footprint of the field at 3.6\,$\mu$m (S--COSMOS; \citealt{Sanders07}).

Overall, the S2COSMOS survey regions provide 1.6 and a further 1.0\,sq.\,degree of 850\,$\mu$m imaging at a median instrumental noise of 1.2 and 1.7\,mJy\,beam$^{-1}$, respectively, and represent a significant improvement in the depth and area coverage of sub--mm imaging of this important survey field. 

\begin{figure}
  \centering
  \includegraphics[width=0.48\textwidth]{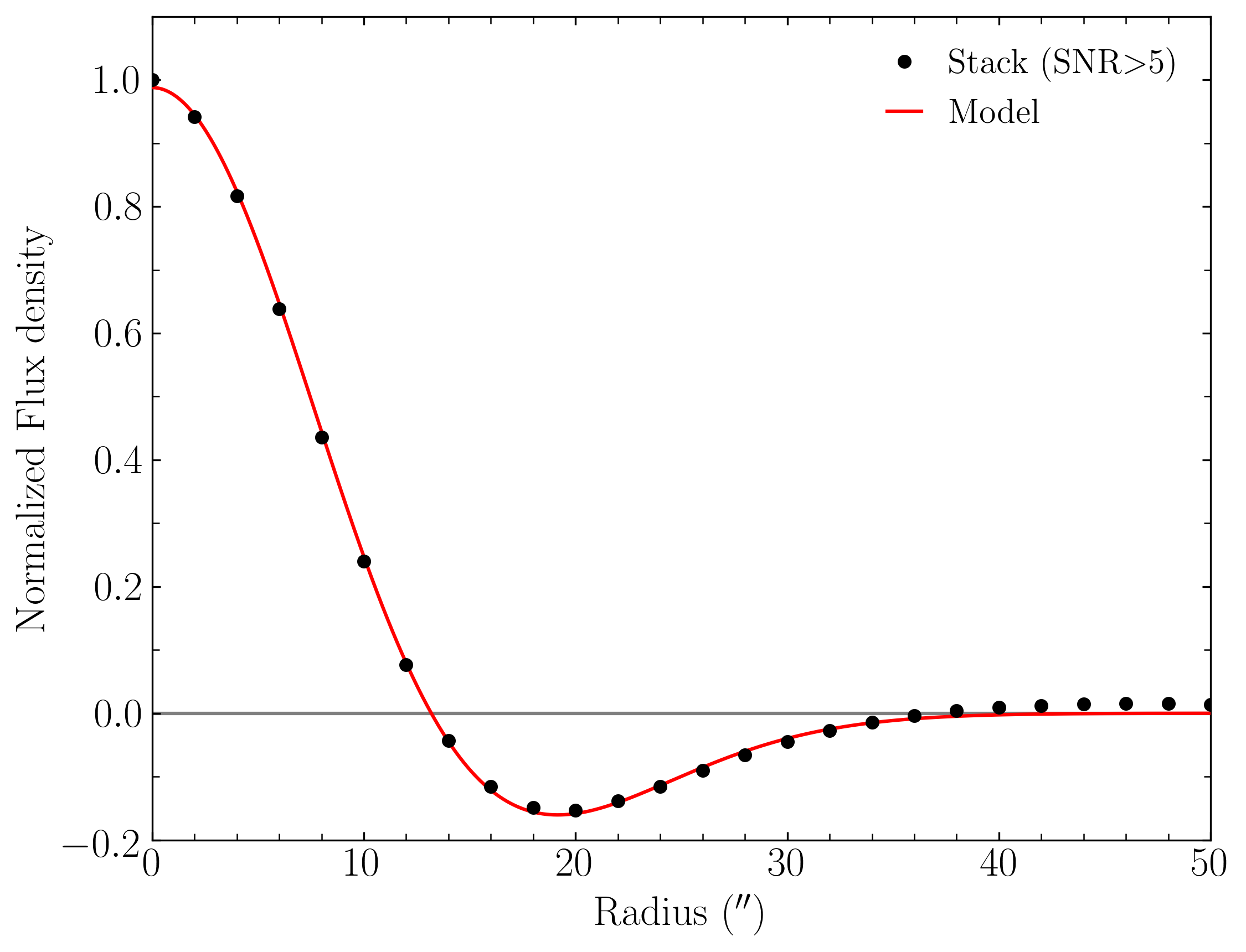} 
  \caption{The normalized, empirical SCUBA--2 PSF at 850\,$\mu$m, created by stacking bright, isolated sources (SNR\,$>$\,5; $>$\,40$''$) in the S2COSMOS map. The negative `ringing' in the PSF is a result of the match--filtering and smoothing applied to the intrinsic SCUBA--2 imaging to enhance sensitivity to point source emission. The empirical PSF is well modeled by the superposition of two Gaussian functions (see also \citealt{Geach17}) and has a FWHM\,=\,14.9$''$.
  }
\label{fig:beamprofile}
\end{figure}

\subsubsection{Beam Profile}
\label{subsubsec:beamprofile}
The response of SCUBA--2\,/\,JCMT to point source emission at 850\,$\mu$m is well-described by the superposition of two Gaussian functions, where the primary (secondary) component has a FWHM\,=\,13$''$ (48$''$) and contains 98\,\pc\ (2\,$\pc$) of the total flux \citep{Dempsey13}. However, we apply a number of filtering steps during the S2COSMOS data reduction that modify the point spread response function (PSF). To determine the effective SCUBA--2\,/\,JCMT PSF, after filtering, we stack the S2COSMOS map at the position of all 850\,$\mu$m sources that are detected at $>$\,5\,$\sigma$ (see \S~\ref{subsec:sourceext}) and that are separated by $>$40$''$. 

The resulting radially-averaged, normalized, stacked profile of the PSF at 850\,$\mu$m is shown in Figure~\ref{fig:beamprofile}. The core of the empirical PSF has a {\sc fwhm}\,=\,14.9$''$ and displays negative ringing that arises due to the matched--filter applied to the map ($15$\,$\pc$ of the normalized peak at a radius of $\sim$\,20$''$). The radially-averaged profile is well--described by the superposition of two Gaussian functions (e.g.\ \citealt{Geach17})

\begin{equation}
G(\theta) = A{_1}\,\mathrm{exp}\left( - \frac{\theta^2}{2\sigma_{1}^2}\right) +  A{_2}\,\mathrm{exp}\left(-\frac{\theta^2}{2\sigma_{2}^2}\right),
\end{equation}

\noindent and we derive best--fit values of $A_{1}$\,=\,3.46, $A_{2}$\,=\,$-$2.46, $\sigma_{1}$\,=\,8.97$''$ and $\sigma_{2}$\,=\,10.82$''$.

\subsubsection{Astrometry}
\label{subsubsec:ast}
Regular observations of standard calibrators are performed during nightly observations with the JCMT to identify, and correct, for large-scale drifts in the telescope pointing. To verify the accuracy of the resulting astrometric solution for the S2COSMOS map we use a reference catalogue of sources detected in observations with the VLA at 3\,GHz \citep{Smolcic17}, leveraging the correlation, at a fixed redshift, between emission at far--infrared and radio wavelengths for star-forming galaxies (e.g.\ \citealt{Yun01}), to obtain a stacked detection of radio sources in the SCUBA--2 map. We stack the S2COSMOS 850\,$\mu$m image at the position of 8850 sources that are detected at a significance level of $>$5.5\,$\sigma$ in the 3\,GHz image (estimated false detection rate of 0.4\,$\pc$), and obtain a strong detection at a SNR\,=\,90\,$\sigma$. The stacked emission is well-centered at the position of the 3\,GHz sources; modeling the stacked emission with the best-fit PSF presented in \S~\ref{subsubsec:beamprofile} we determine small, but statistically--insignificant, offsets of $\Delta$\,R.A.\,=\,$-0.2$\,$\pm$\,0.1$''$ and $\Delta$\,Dec.\,=\,0.1\,$\pm$\,0.1$''$. Thus, as the astrometry of the S2COSMOS and 3\,GHz\,/\,VLA maps are in such close agreement we do not apply any systematic corrections to our 850\,$\mu$m imaging.

\section{Analysis}
\label{sec:analysis}
\begin{figure}
  \includegraphics[width=0.48\textwidth]{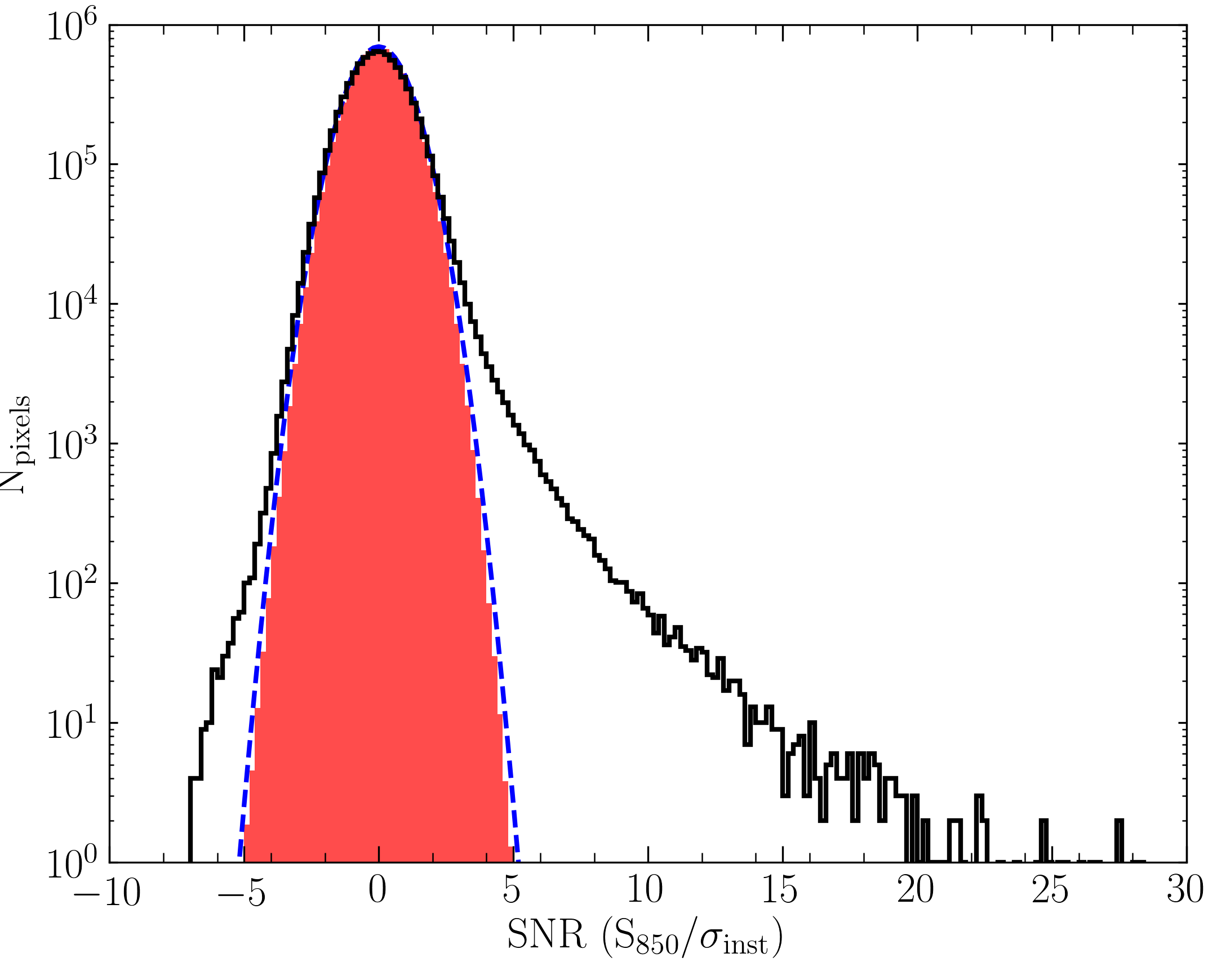}
  \caption{ The signal--to--noise ratio distribution (solid line) for pixels in the total 2.6\,sq.\,degree S2COSMOS survey area ($\sigsubm$\,$<$\,3\,\mjpb). The shaded region represents the average of 40 jackknife maps that were created by randomly inverting the flux densities and co-adding half of the observations, and demonstrates that the instrumental noise is Gaussian in nature (dashed line; mean of zero and standard deviation of one). The data distribution show an excess of positive and negative emission relative to the jackknife distributions that represent astrophysical sources and the effect of match-filtering, respectively. 
  }
\label{fig:snrdist}
\end{figure}

The S2COSMOS survey provides 2.6 sq.\,degree of 850$\mu$m imaging at an instrumental noise level of 0.5--3.0\,mJy\,beam$^{-1}$. In Figure~\ref{fig:snrdist} we show the histogram of pixel signal--to--noise ratio across the {\sc main} and {\sc supp} regions. The signal--to--noise ratio histogram for the S2COSMOS survey displays three clear features; a strong tail of positive emission that extends to a SNR\,=\,30 and represents real astrophysical emission; a central region that is broadly consistent with Gaussian noise; and, excess negative emission that arises due to the negative ringing around positive emission that is introduced in the match-filtering step. The aim of our survey is to extract the position and flux density of astrophysical sources that are detected in the S2COSMOS image and we discuss that process here.

\subsection{Source Extraction}
\label{subsec:sourceext}
By applying a matched-filter to the S2COSMOS map we have optimized the image for the detection of point source emission in the presence of instrumental noise. To extract sources from the S2COSMOS image we thus use a ``top-down'' approach to sequentially identify and subtract the highest significance sources detected across the map. First, the highest signal--to--noise ratio pixel in the S2COSMOS image is identified, and the flux density and position of the source is recorded. Next, the emission is modeled using the empirical PSF derived in \S~\ref{subsubsec:beamprofile} and the best-fit for this source is subtracted from the image. If a source is identified within $40''$ of a prior detection then we account for the potential blending of the emission by re-injecting the nearest source into the map and modeling the emission with a double PSF model. The process of isolating and removing sources of emission is repeated until a floor--threshold at 3.5\,$\sigma$ is reached, with all sources detected above this significance level recorded in a preliminary catalogue.

To construct a robust catalogue of 850\,$\mu$m sources for further analysis we require knowledge of the ratio of spurious to total detections across the S2COSMOS map, the false--detection--rate (FDR). We estimate the FDR for our survey using 40 jackknife realizations of the S2COSMOS map. Each jackknife realization is created by randomly inverting half of the flux densities of individual SCUBA--2 scans, separated by scan pattern and pointing centre, before co-adding and match-filtering the resulting map. The jackknife process removes any sources of astrophysical emission and the resulting maps provide realistic realizations of the instrumental noise profile. We apply our source extraction procedure to the jackknife maps and catalogue any ``sources'' in an identical manner to our preliminary source catalogue.

\begin{figure}
  \includegraphics[width=0.48\textwidth]{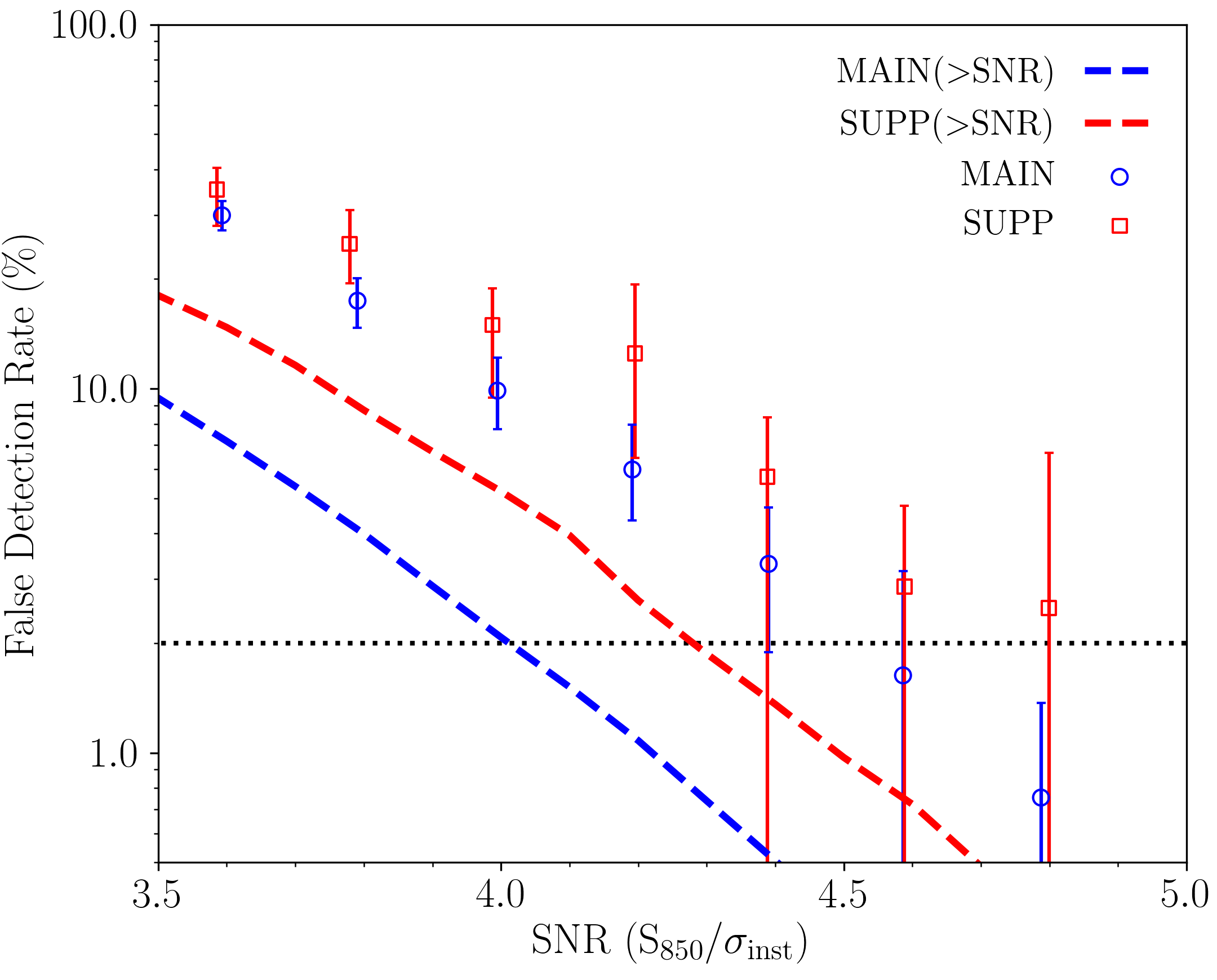}
  \caption{Using our jackknife maps we estimate the false--detection rate (defined as the ratio of `detected' sources in jackknife and observed maps) as a function of detection significance in the S2COSMOS {\sc main} and {\sc supp} regions. The integrated FDR across each catalog is represented by a dashed line, while data points show the FDR at a specific SNR. At a 4\,$\sigma$ threshold for detection we estimate that our {\sc main} catalog has an integrated FDR of 2\,\% (dotted line) and we adopt that threshold here. At fixed SNR the FDR rate increases in the {\sc supp} region relative to the {\sc main} survey, reflecting the lower sensitivity and steep slope in the bright--end of the 850\,$\mu$m number counts. To ensure that the S2COSMOS catalog provides a robust sample for future study we adopt a 4.3\,$\sigma$ threshold for detection in the {\sc supp} region, corresponding to a FDR\,=\,2\% across the entire S2COSMOS source catalog.}
\label{fig:snrandfdr}
\end{figure}

Using the catalog of sources that are detected in the S2COSMOS image and the jackknife maps we construct the FDR of our image as function of signal-to-noise ratio (Figure~\ref{fig:snrandfdr}). The integrated FDR within the S2COSMOS {\sc main} survey region is 2\,$\pc$ at $>$\,4\,$\sigma$, and we adopt this as the detection limit throughout our analysis. The FDR is estimated to be higher in the {\sc supp} area, at a fixed signal--to--noise ratio, reflecting the lower sensitivity achieved in this region and the steep slope of the 850\,$\mu$m number counts. To account for this increasing FDR we apply a $>$\,4.3\,$\sigma$ threshold for detection within the {\sc supp} region, at which we estimate that our {\sc supp} catalog has a spurious fraction of 2\,$\pc$, consistent with our {\sc main} sample.

At our detection limits of $\ge$\,4\,$\sigma$ and $\ge$\,4.3\,$\sigma$ we identify 1020 and 127 bright 850\,$\mu$m sources that are located within the S2COSMOS {\sc main} and {\sc supp} regions, respectively. Based on the expected FDR we estimate that 21 and 2 sources in the {\sc main} and {\sc supp} catalog are spurious, respectively. The S2COSMOS source catalogue presented here contains 1147 sub--mm sources with observed 850\,$\mu$m flux densities from 2--20\,mJy, providing a uniquely large sample with which to study the properties of intensely star-forming, dust-obscured galaxies and their relation to other galaxy populations in the COSMOS field.

\subsection{Flux boosting and Completeness}
\label{subsec:completeness}
To test the efficiency of our source extraction we create simulated maps of the S2COSMOS footprint. These simulations are important to determine two key aspects of our survey: the completeness as a function of intrinsic flux density; and the accuracy of the measured flux density and associated uncertainty for each source in the S2COSMOS catalogue. 

It is well-known that the flux density of a source in a signal--to--noise limited catalogue will be biased if the source counts are non-uniform. The effect is related to Eddington bias and, at sub-mm wavelengths, where the bright--end of the source counts are steep \citep{Scott02,Karim13,Simpson15b,Geach17}, the effect is commonly referred to as flux boosting. This nomenclature reflects that there is a higher probability that a source of a given flux density corresponds to a fainter source that is scattered upwards in flux density, due to Gaussian noise fluctuations, than a brighter source that is scattered downwards. Thus, at a fixed signal--to--noise ratio a source appears brighter on average, although the magnitude of the boosting is both a function of the local noise and the intrinsic flux density of the source.

Both Bayesian and empirical approaches have been adopted to characterize the effect of flux boosting on surveys at sub-mm wavelengths (e.g.\ \citealt{Coppin06,Casey13}). However, regardless of the method that is adopted these techniques require an input model for the intrinsic source counts of the underlying population that imprints prior information on the results. In this work we adopt an empirical approach to determine the effect of flux boosting, but rather than assuming a prior estimate for the intrinsic number counts we first iterate towards an input model that broadly reproduces the observed distribution of flux densities for the S2COSMOS source catalogue (e.g.\ \citealt{Wang17}).

\begin{figure*}
  \centering{\includegraphics[width=\textwidth]{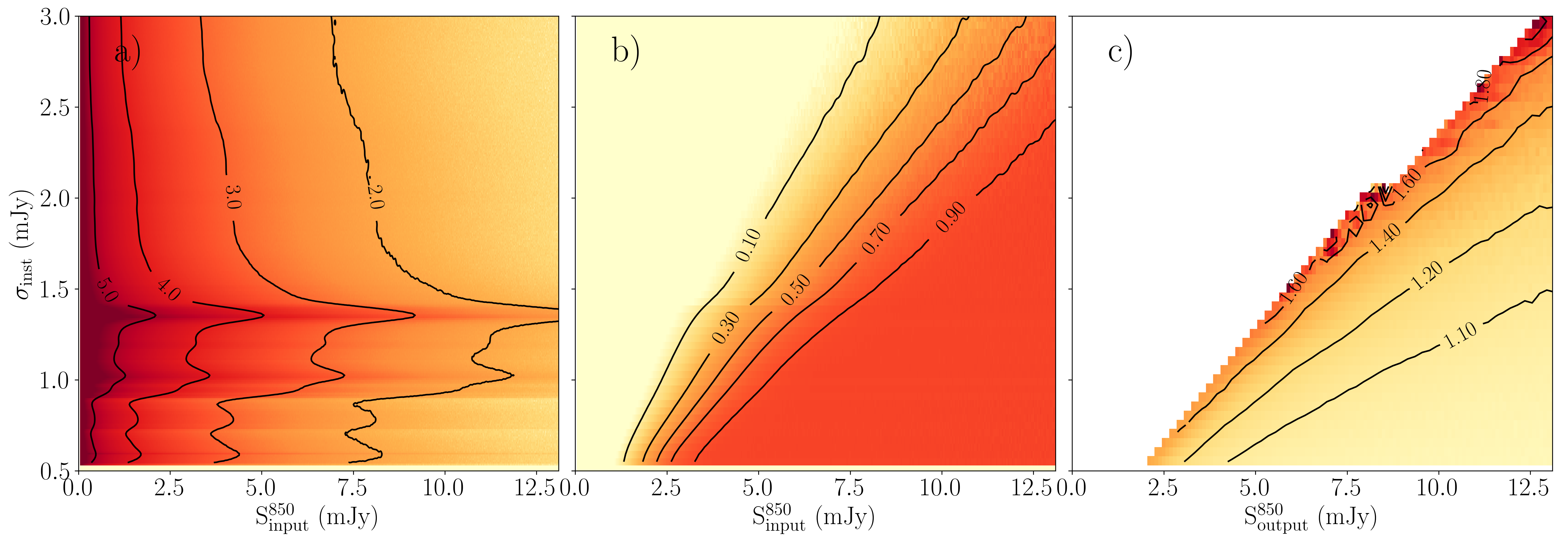}}
  \caption{ We create 10$^{5}$ simulations of the S2COSMOS survey to determine the effect of flux boosting and our survey completeness. Here we present the results of these simulations showing: a) the number of injected sources (contours labelled in log$_{10}(N)$) at a given input flux density and instrumental sensitivity; b) completeness to source of a given input flux density and instrumental sensitivity; and c) the ratio of output--to--input flux density (flux boosting) as a function of measured flux density and instrumental sensitivity. Note that the structure in panel a reflects variation in the S2COSMOS noise map, and the discontinuity in panel c corresponds to the change in detection threshold from SNR\,$>$\,4 and SNR\,$>$\,4.3 {\sc main} and {\sc supp} regions, respectively. The density peaks in the number of injected sources reflect variations in the instrumental noise level of the S2COSMOS map. Within our {\sc main} survey area the overall completeness is 50\,\% (90\,\%) for sources with an intrinsic flux density of $\fluxsubmint$\,=\,4.4\,mJy (6.4\,mJy), or $\fluxsubmint$\,=\,5.1\,mJy (9.1\,mJy) when considering the {\sc main}\,+\,{\sc supp} region. As expected, the magnitude of flux boosting is function of both flux density and instrumental noise, with the average correction reaching a factor of $\sim$\,1.8 in the outskirts of our survey area. In practice, sources detected in the S2COSMOS map are deboosted based on both their local instrumental noise and measured flux density.}
\label{fig:numcompdeb}
\end{figure*}

To estimate the shape of the intrinsic 850\,$\mu$m number counts we use a set of source simulations that are designed to produce realistic mock versions of the S2COSMOS map and source catalog. First, we adopt the best--fit 850\,$\mu$m number counts presented by \citet{Geach17} to provide a plausible, starting estimate for the shape of the intrinsic counts. Next, a jackknife realization of the S2COSMOS survey is chosen at random, and simulated sources are injected into the map down to a flux density limit of 0.05\,mJy, following the shape and normalization of the input number counts. Each source is placed at a random position in the jackknife map and is injected based on the empirical PSF constructed in \S\,\ref{subsubsec:beamprofile}. We note that the clustering strength of 850\,$\mu$m sources, especially as a function of redshift and luminosity, is not currently well constrained and as such we do not include any contribution from this effect in our simulations (\citealt{Hickox12,Wilkinson17}). The process is repeated to create 100 simulated maps of the S2COSMOS survey and sources are extracted from these simulated maps in the same manner as for the ``true'' observations. Finally, we use our catalog of extracted, simulated sources to construct the observed differential number counts and compare these to the raw counts for the S2COSMOS survey.

To improve our estimate of the intrinsic number counts we consider the measured offset between each bin in the simulated and observed number counts. However, to apply these offsets as a correction to the input model we must account for the fact that each bin in the simulated counts is comprised of sources that have a range of intrinsic flux densities. As such, we first map each source that contributes to the simulated counts to a bin in the intrinsic flux distribution of all sources that were injected into the simulated map, and store the relevant offset from the comparison of the observed and simulated counts. Note that we consider a source recovered in the simulation if it is the brightest component within 11$''$ of a detected source (radius\,=\,0.75\,$\times$\,{\sc{fwhm}}). Finally, the average correction is applied to each bin in the intrinsic distribution of injected source and these are modeled with a \citet{Schechter76} function of the form

\begin{equation}\label{eqn:Schechter}
\frac{dN}{dS} = \frac{N_{0}}{S_0} \left( \frac{S}{S_0}\right)^{-\gamma} \mathrm{exp}\left( \frac{-S}{S_0} \right),
\end{equation}

and the best-fit values of $N_{0}$, $S_{0}$, and $\gamma$ are used as the input model in the next iteration. This procedure is repeated for twenty ``major'' iterations and the process rapidly converges towards an input count model with $N_{0}$\,$\sim$\,5300\,deg$^{-2}$, $S_{0}$\,$\sim$\,2.9\,mJy, and $\gamma$\,$\sim$\,1.5.

The simulations described above provide a first-order approximation of the intrinsic 850\,$\mu$m number counts. To derive accurate flux boosting and completeness corrections on a source-by-source basis we now create 10$^{5}$ simulations of the S2COSMOS image using the best--fit Schechter function described above as the input model for the 850\,$\mu$m number counts. The result of the source simulations are shown in Figure~\ref{fig:numcompdeb}, where we present the number of injected sources and the completeness as a function of both instrumental noise and input flux density, and the effect of flux boosting as function of instrumental noise and {\it observed} flux density. 

From our source simulations we estimate that the S2COSMOS {\sc main} and {\sc main}\,$+$\,{\sc supp} catalogs achieve 50\,$\pc$ (90\,$\pc$) completeness at an intrinsic flux density of 4.4\,mJy (6.4\,mJy) and 5.1\,mJy (9.1\,mJy), respectively. These completeness levels reflect the integrated completeness across the survey regions, taking into account variation in the noise level. Comparing the ratio of input to output flux density of the recovered source we estimate that the flux density of a source in our {\sc main} survey area is boosted on average by 55\,$\pc$ and 6\,$\pc$ for a detection significance of 4\,$\sigma$ and 10\,$\sigma$, respectively. The effect of flux boosting is expected to be a function of both the observed flux density and local noise and this dependence is evident in our simulation (see Figure~\ref{fig:numcompdeb}). In the deepest regions of the S2COSMOS map ($\sigma^{\mathrm{inst}}_{850}$\,$<$\,0.7\,mJy\,beam$^{-1}$) the flux density of a source that is identified at a SNR\,=\,4 is boosted on average by $\sim$\,40\,$\pc$, increasing to $\sim$\,55\,$\pc$ for a source identified at the detection threshold in the higher noise {\sc supp} region (SNR\,=\,4.3). Thus, in practice each S2COSMOS source is deboosted based on its local noise estimate and observed flux density. For a given value of the local noise and observed flux density the distribution of true flux densities is constructed using the results of our simulations. The median and the 16--84$^{\mathrm{th}}$ percentile range of the resulting distribution is taken as the deboosted flux density and its associated uncertainty. Table~1 lists the deboosted flux densities and associated uncertainties for each source in the S2COSMOS catalogue and we use these deboosted values when considering the flux density of a source in the remainder of our analysis.

Finally, the source simulations provide an estimate of the positional uncertainty associated with each S2COSMOS source. From the catalogue of simulated sources we calculate the angular offset between the injected position and the recovered position of each source. We estimate a median uncertainty of $\sim$\,3$''$ on the radial position of sources that are detected at the 4\,$\sigma$ significance level, with 95\,$\pc$\, of sources offset by $<$\,8.7$''$.

\subsection{Confusion noise}
\label{subsec:confusion}
Next we consider the effect of confusion noise, arising due to the blending of faint galaxies within the JCMT beam, on the properties of the S2COSMOS image. By the standard `rule of thumb' the confusion limit of an image is reached when the surface density of sources reaches one per $\sim$\,20--30 resolution elements (e.g.\ \citealt{Condon74,Hogg01,Takeuchi04}). Adopting this criterion we estimate that the S2COSMOS image has a confusion limit of $\sim$\,2\,mJy, or $\sigma_{\mathrm c}$\,$\sim$\,0.5\,mJy at our 4\,$\sigma$ threshold for detection in the {\sc main} S2COSMOS survey. 

Our simple estimate for the confusion noise does not account for the properties of the S2COSMOS map and our source extraction procedure, and is sensitive to the underlying distribution of source flux densities (see \citealt{Takeuchi04}). To provide a more realistic estimate of the JCMT\,/\,850\,$\mu$m confusion noise we next consider the properties of the S2COSMOS map and the results of our extensive source simulations (see \S\,\ref{subsec:completeness}). Following \citet{Dole03}, the photometric confusion limit can be defined by the standard deviation of beam--to--beam fluctuations ($\sigma_{\mathrm{c}}$) below a limiting flux ($S_{\mathrm{lim}}$), where $S_{\mathrm{lim}}$\,=\,$q$\,$\sigma_{\mathrm{c}}$ and we assume $q$\,=\,4 to match the adopted significance threshold for detection within the S2COMSOS {\sc main} survey region (see also \citealt{Dole04,Frayer06,Frayer09,Nguyen10,Magnelli13}). Adopting an upper limit ($S_{\mathrm{lim}}$) when estimating the beam--to--beam fluctuations ensures that the brightest sources at 850\,$\mu$m do not skew any estimate of the confusion noise to a high, potentially unbounded, value (see \citealt{Valiante16}). To estimate the confusion noise inherent on the S2COSMOS image we adopt an iterative approach that is based on the source extraction procedure described in \S~\ref{subsec:sourceext}. First, an upper limit to the confusion noise is estimated following

\begin{equation}
\sigma_{\mathrm{c}} = \sqrt{\sigma_{\mathrm{total}}^2 - \sigma_{\mathrm{inst}}^2},
\label{eqn:conf}
\end{equation}

\input{tableshort.tex}

\noindent where $\sigma_{\mathrm{total}}$ and $\sigma_{\mathrm{inst}}$ represent the standard deviation of the S2COSMOS 850\,$\mu$m map and jackknife image, respectively. The instrumental noise is expected to dominate over confusion for the majority of the S2COSMOS image and, as such, we only consider the deepest 0.1 sq.\ degree region of the S2COSMOS map at $\sigma^{\mathrm{inst}}_{850}$\,$<$\,0.7\,mJy\,beam$^{-1}$ in our analysis. Next, we identify the highest significance detection across the S2COSMOS image and, if the flux density of the source is greater than $S_{\mathrm{lim}}$, the best--fit model is subtracted from the image. Finally, $\sigma_{\mathrm{total}}$ is calculated from the residual, source--subtracted image and the confusion noise is re-evaluated following Equation~\ref{eqn:conf}. The source identification and extraction process is repeated until the confusion noise converges at $\sigma_{\mathrm{c}}$\,=\,0.34\,mJy\,beam$^{-1}$, at the S2COSMOS threshold for detection (SNR\,=\,4). Note that if we consider the 2.6 sq.\ degree S2COSMOS {\sc main} and {\sc supp} survey region then we estimate $\sigma_{\mathrm{c}}$\,=\,0.50\,mJy\,beam$^{-1}$, reflecting the contribution to the total noise that arises from sources that lie below the threshold for detection but above the true confusion limit.

Next, we use the results of our source simulations to provide a further estimate of the confusion noise on the S2COSMOS image. From our catalog of injected and extracted model sources we construct the distribution of measured source flux densities as a function of input flux density and local instrumental noise. The width of the measured flux density distribution represents the total uncertainty due to instrumental noise and source confusion. Again, we consider sources that are injected within the 0.1 sq.\ degree, $\sigma^{\mathrm{inst}}_{850}$\,$<$\,0.7\,mJy\,beam$^{-1}$ region of the simulated S2COSMOS map and limit our analysis to input flux densities where the source catalog is 95\,$\pc$ complete (see \S~\ref{subsec:completeness}; SNR\,$=$\,6). The total noise ($\sigma_{\mathrm{total}}$) is estimated from the 16--84$^{\mathrm{th}}$ percentile of the distribution of measured flux densities and, following Equation\,\ref{eqn:conf}, we estimate a confusion noise of $\sigma_{\mathrm{c}}$\,=\,0.36\,$\pm$\,0.02\,mJy\,beam$^{-1}$. Note that if we require that the measured flux density distribution is 99\,$\pc$ complete then the estimate for the confusion noise increases to $\sigma_{\mathrm{c}}$\,=\,0.42\,$\pm$\,0.02\,mJy\,beam$^{-1}$.

Overall, we conclude that the confusion noise on the S2COSMOS image is $\sigma_{\mathrm{c}}$\,$\sim$\,0.4\,mJy\,beam$^{-1}$, in agreement with previous estimates from ``pencil--beam'' ($\lsim$\,0.02 sq.\ degree), confusion--dominated SCUBA--2 imaging at 850\,$\mu$m (\citealt{Zavala17,Cowie17}). Importantly, the instrumental noise dominates across the {\sc main} S2COSMOS survey region (median $\sigma^{\mathrm{inst}}_{850}$\,=\,1.2\,mJy\,beam$^{-1}$), and only approaches the confusion noise in the deepest regions of the map (0.05\,sq.\,degree at median $\sigsubm$\,=\,0.6\,mJy\,beam$^{-1}$). As such, we do not consider the effect of confusion noise on the S2COSMOS survey in further detail. Note that the effects of source confusion are inherent in our simulated maps of the S2COSMOS survey and, as such, the associated uncertainty on the deboosted flux density of each S2COSMOS source includes a contribution from both confusion and instrumental noise. 

\section{Discussion}
We have presented the deep, 850\,$\mu$m imaging and source catalog for the S2COSMOS survey. Across the 2.6 sq.\,degree of the full S2COSMOS field we detect 1147 sub-mm sources with intrinsic flux densities of S$_{850}$\,=\,2--20\,mJy. We now present a discussion of the fundamental 850\,$\mu$m properties of the galaxies that are covered by the S2COSMOS imaging. Initially we focus on the properties of the highest--luminosity, individually--detected sources (\S~4.1--4.3), which comprise each of our source catalogs, before presenting a stacking analysis of lower--luminosity, mass-selected samples (\S\,4.4).

\subsection{Number Counts}
The number of detected 850\,$\mu$m sources as a function of flux density is a fundamental output from our large area and contiguous survey. The submm number counts can provide a powerful, simple test of models of galaxy formation that is free from further physical interpretation of the observed quantities (e.g.\ \citealt{Baugh05}). Furthermore, studying the variation in the number counts that are constructed from observations of different survey fields, resulting from cosmic variance, can in principle provide insights into the underlying properties of the galaxy population. Indeed, determining whether the 850\,$\mu$m number counts are strongly affected by cosmic variance is the first step to testing if sub-mm sources are, as is often suggested, a highly--biased tracer of the underlying matter distribution of the Universe (e.g.\ \citealt{Scott02,Blain04a,Chapman09,Hickox12}; but see also \citealt{Danielson17,Wilkinson17}), and determines whether our survey is sufficiently large to be a fair representation of the underlying source population.

To determine the number counts at 850\,$\mu$m we consider the 1020 and 127 sources that are detected at SNR\,$>$\,4 and SNR\,$>$\,4.3 across the S2COSMOS {\sc main} and {\sc supp} regions, respectively. For both the {\sc main} and {\sc main+supp} region, the differential and cumulative counts are constructed using the deboosted flux density for each S2COSMOS source, with completeness corrections calculated and applied based on the results of injecting simulated sources into the S2COSMOS jackknife maps (see \S\,\ref{subsec:completeness}). The associated uncertainty on the deboosting correction for each source can be significant and, crucially, follows a non-Gaussian distribution. To ensure that our measurement of the number counts captures this information we construct 10$^{4}$ realizations of the S2COSMOS source catalog. In each realization we assign a deboosted flux density to a source by randomly sampling from the full distribution of possible intrinsic values based on the observed flux and local noise level of the original detection. The counts are constructed from each realization and the median and 16--84th percentile of the resulting distribution are taken as the final number counts and associated uncertainties for both the {\sc main} and {\sc main+supp} regions (see Table~2). 

\begin{figure*}
  \centering\includegraphics[width=0.65\textwidth]{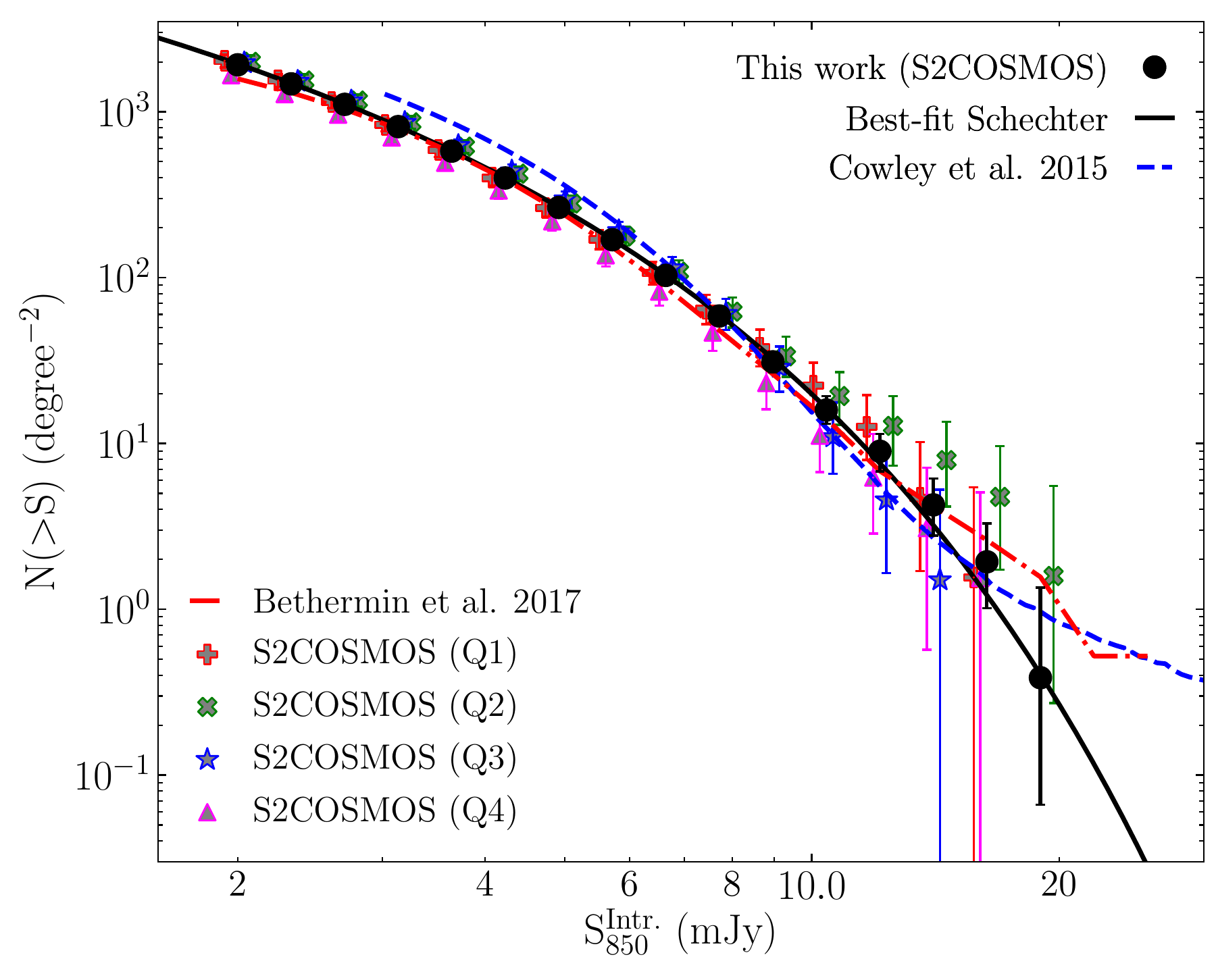}
  \hfill
  \includegraphics[width=0.65\textwidth]{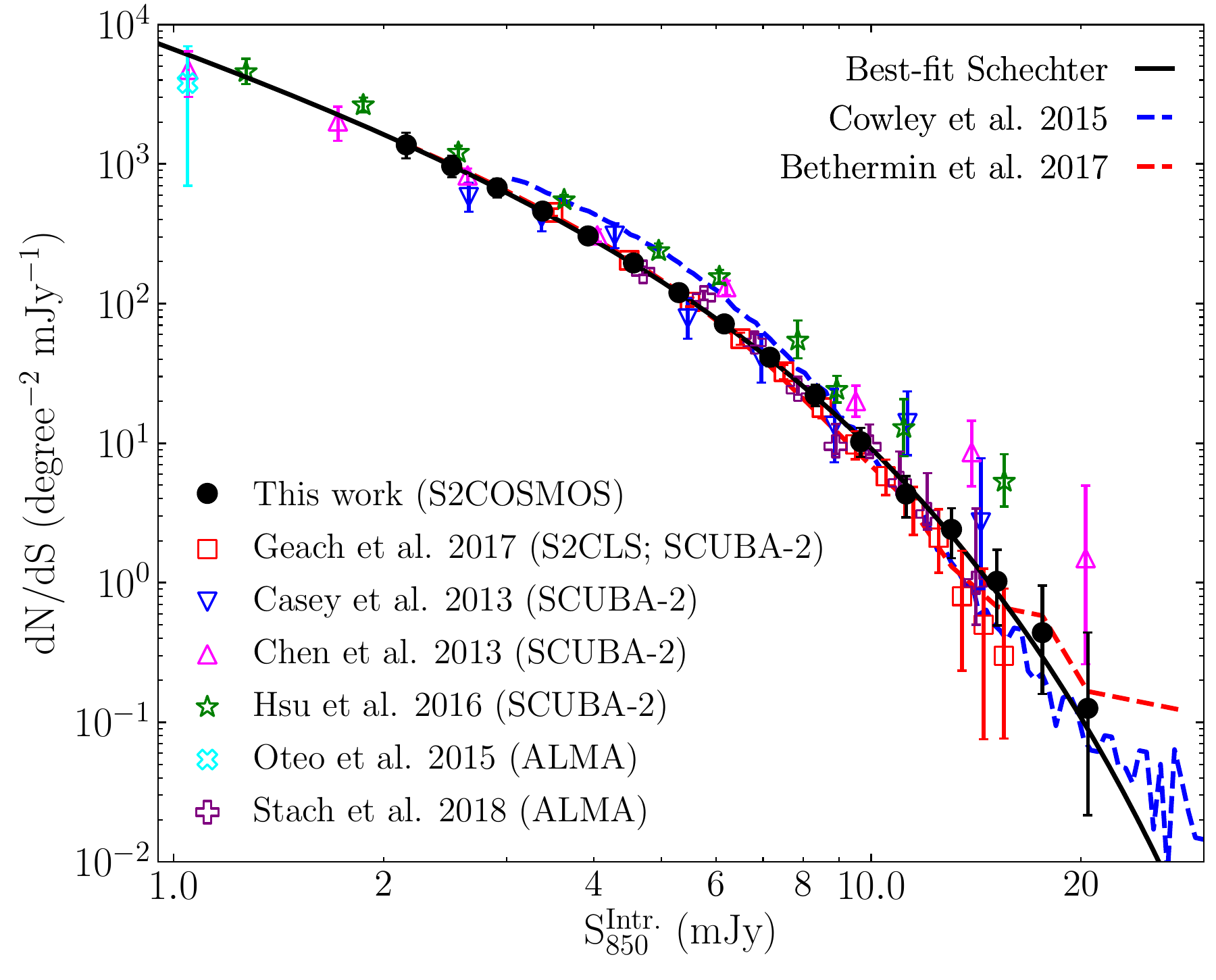}
  \caption{The differential and cumulative 850\,$\mu$m number counts constructed from the S2COSMOS survey (covering S$_{850}$\,=\,2--20\,mJy), compared to previous SCUBA--2 surveys and theoretical predictions. {\textit{Top:}} The cumulative S2COSMOS number counts constructed from the full 2.6 sq.\ degree survey region are shown, as well as those constructed from four contiguous, independent 0.65 sq.\ degree regions of our 850\,$\mu$m imaging ({\sc main} and {\sc supp}). For clarity the cumulative counts constructed from each quadrant are displayed with an small offset in flux density. Overall, the cumulative counts constructed from each quadrant are in good agreement with those constructed from our full survey region and any scatter at the bright--end ($S_{850}$\,$>$\,10\,mJy) is consistent within the associated uncertainties. The full--survey S2COSMOS cumulative counts are well-fit by a single Schechter function with no evidence for a deviation at high flux densities, indicating that our sample suffers minimal contamination from low--redshift\,/\,galactic ($z$\,$\lsim$\,0.1) or strongly--lensed sources, although we note that we cannot rule out the contribution from weak lensing (e.g.\ \citealt{Almaini05,Aretxaga11,Bourne14}). {\textit{Bottom:}} The S2COSMOS differential number counts, along with a selection of counts constructed from: previous single--dish surveys of blank--fields \citep{Casey13,Geach17} and lensing clusters (e.g.\ \citealt{Chen13,Hsu16}); ALMA follow--up observations of S2CLS--selected sources in the UDS \citep{Stach18}; and serendipitous detections in ALMA calibrator observations (\citealt{Oteo16a}). For comparison, we show the estimated counts at 850\,$\mu$m from the phenomenological model of \citet{Bethermin17} and the predictions from the semi--analytic model GALFORM \citep{Cowley15}, both of which attempt to model source blending within the JCMT beam. The theoretical models are in broad agreement with our results at $S_{850}$\,$\gsim$\,7\,mJy, while at fainter flux densities the GALFORM model lies $\sim$\,1.4--1.6\,$\times$ above the S2COSMOS counts. }

\label{fig:counts}
\end{figure*}

As discussed in \S\,\ref{sec:analysis}, the S2COSMOS {\sc supp} region provides 1 sq.~degree of shallower 850\,$\mu$m coverage in addition to our deep, 1.6 sq.~degree {\sc main} survey, and increases our area coverage for rare, luminous sources. We have ensured a consistent FDR across both the {\sc main} and {\sc supp} source catalogs, but the higher instrumental noise level in the {\sc supp} region results in typically larger, more uncertain corrections for flux boosting. To investigate whether this increased uncertainty affects our results we compare our estimates of the 850\,$\mu$m number counts that are constructed from the {\sc main} and {\sc main+supp} regions. We identify a small, statistically--insignificant increase of, on average, 2\,$\pm$\,1\,$\pc$ in the differential counts constructed from the {\sc main} region, relative to {\sc main+supp}, and, similarly, no significant change in the cumulative counts. Notably, including sources detected in the {\sc supp} region reduces the associated, fractional uncertainties on our estimate of the 850\,$\mu$m differential counts by an average of 13\,$\pm$\,7\,$\pc$, increasing to 20--70\,$\pc$ at the highest flux densities ($>$\,8\,mJy; Table\,2). Considering the agreement between the number counts constructed from each of our survey regions, and the relative improvement in the associated uncertainties, we choose to adopt the results from {\sc main+supp} survey in the following analysis.

The differential and cumulative 850\,$\mu$m number counts constructed from the S2COSMOS survey ({\sc main+supp}) are shown in Figure~\ref{fig:counts} and presented in Table\,2. As can be seen in Figure~\ref{fig:counts}, the number counts follow a smooth, exponential decline with increasing flux density. At the highest flux densities (S$_{850}$\,$\gsim$\,15\,mJy) it is expected that both low--redshift\,/\,galactic ($z$\,$\lsim$\,0.1) and strongly--lensed sources will start to strongly influence the number counts (e.g.\ \citealt{Negrello10,Vieira10}). The result is an excess in the number counts relative to an exponential decline that has been confirmed by wide--area surveys with {\it Herschel} at 500\,$\mu$m \citep{Negrello10,Wardlow13,Valiante16} and the SPT at 1.4\,mm \citep{Vieira10}. Note that gravitationally--lensed source are expected to contaminate the counts at lower flux densities (S$_{850}$\,$\lsim$\,15\,mJy; e.g.\ \citealt{Bourne14}), subtly changing the shape of the expected exponential decline, but this is not expected to be a dominant effect and requires robust identifications for each S2COSMOS source to quantify (e.g.\ \citealt{Simpson17}). We investigate the S2COSMOS number counts and find that at the brightest flux densities they do not show any evidence for such an excess, with the brightest source in our survey identified at $S_{850}$\,=\,19.8$^{+1.6}_{-2.0}$\,mJy ({\sc main} sample, but not located in S2CLS-COSMOS coverage; \citealt{Geach17}). The absence of an excess in the S2COSMOS counts is statistically consistent with the results from the S2CLS survey, which identified three sources at S$_{850}$\,$\gsim$\,20\,mJy over $\sim$\,4 sq.\,degree and a mild excess in the number counts. Considering both S2COSMOS and S2CLS we conclude that any enhancement in the bright 850\,$\mu$m counts due to low--redshift\,/\,galactic ($z$\,$\lsim$\,0.1) and strongly--lensed sources is minimal, and subject to low number statistics, on scales of $\lsim$\,5\,sq.\,degree.

Due to the lack of any observed excess at bright flux densities we model the S2COSMOS differential number counts with a single Schechter function (Equation~\ref{eqn:Schechter}), determining best-fit parameters of $N_{0}$\,=\,5000$^{+1300}_{-1400}$\,deg$^{-2}$, $S_{0}$\,=\,3.0$^{+0.6}_{-0.5}$\,mJy, and $\gamma$\,=\,1.6$^{+0.3}_{-0.4}$. The best-fit values are in close agreement with the input model used in our deboosting simulations, confirming the strong internal consistency of our source--by--source deboosting corrections (see Table~1). In Figure~\ref{fig:counts} we compare the S2COSMOS number counts to previous surveys at 850\,$\mu$m. Overall, the measured S2COSMOS differential number counts are in reasonable agreement with the results of previous studies, where these directly overlap in flux density. For brevity, we focus on a direct comparison between the S2COSMOS number counts and the results from S2CLS, the largest--area survey that has been conducted at 850\,$\mu$m. To allow an accurate comparison we repeat our analysis to derive the S2COSMOS number counts in flux density bins that are matched to the results from S2CLS (\citealt{Geach17}). Overall, the S2COSMOS and S2CLS differential counts are found to be in excellent agreement, on a bin-by-bin basis, and any differences are measured at the $<$\,1\,$\sigma$ significance level (Figure~\ref{fig:counts}). At the faint--end an extrapolation of our best-fit model is consistent with deep, small--area studies of lensing clusters \citep{Chen13,Hsu16}. Comparing to number count estimates from ALMA imaging at 870\,$\mu$m (assuming flux density scales as $\nu^{2}$) we find that the S2COSMOS counts are in good agreement with those estimated by \citet{Stach18} from a follow--up survey of SCUBA--2 sources at $S_{850}$\,$>$\,4\,mJy in the UDS field (AS2UDS; normalization is 1.06\,$\pm$\,0.08\,$\times$ lower at $>$\,4\,mJy, relative to S2COSMOS), and those presented by \citet{Oteo16a} at 870\,$\mu$m, although the latter of these have significant associated uncertainties. 

\subsection{Cosmic variance}
If SMGs represent a biased tracer of the underlying matter distribution of the Universe then we can expect that this will manifest as variance in the counts in excess of Poisson noise. Using our large area and homogeneous survey we now investigate the effect of cosmic variance on the 850\,$\mu$m source counts. First, we sub--divide the S2COSMOS survey into four contiguous, independent quadrants. The regions are chosen to ensure that each quadrant provides coverage over $\sim$\,0.65 sq.~degree with a broadly comparable noise profile. Next, we identify sources that are detected in each quadrant and construct the number counts in an identical manner to the overall S2COSMOS survey. As can be seen in Figure\,\ref{fig:counts}, the cumulative number counts constructed from each quadrant are in close agreement with the overall S2COSMOS counts. Considering flux densities $>$\,3\,mJy, we find that the cumulative counts in three of the four quadrants are within 1--$\sigma$ of the combined S2COSMOS counts, with the counts constructed from the remaining quadrant (bottom left; Figure~1) offset at the 1.7\,$\sigma$ significance level at $<$\,6\,mJy.

The level of agreement between the 850\,$\mu$m number counts on scales of $\sim$\,0.65 sq.~degree is consistent with the results from the S2CLS survey. Indeed, as demonstrated by \citet{Geach17}, of the seven S2CLS survey fields only the counts constructed from the 0.1\,sq~degree imaging of the GOODS--N field show a modest (2\,$\sigma$) enhancement relative to the overall S2CLS counts. However, by comparing the number counts derived from the full S2COSMOS survey with those from S2CLS we can extend our analysis to investigate whether cosmic variance affects the 850\,$\mu$m number counts on scales of up to $\sim$\,3~sq.\,degree. As we have demonstrated, each bin in the differential number counts from S2COSMOS and S2CLS are in close agreement, but agreement in each flux bin of the differential counts can mask a significant difference in the integrated number density of sources. Thus, we integrate the differential counts from S2CLS and compare the cumulative number counts to the results presented here. We find excellent agreement in the S2COSMOS and S2CLS cumulative counts at S$_{850}$\,$>$\,3\,mJy (N$^{\mathrm{S2COSMOS}}$\,/\,N$^{\mathrm{S2CLS}}$\,=\,1.01\,$\pm$\,0.05) and a small, but statistically--insignificant excess in S2COSMOS at S$_{850}$\,$>$\,8\,mJy (N$^{\mathrm{S2COSMOS}}$\,/\,N$^{\mathrm{S2CLS}}$=\,1.2\,$\pm$\,0.2), confirming the overall excellent agreement between the number counts constructed from the two surveys.

\input{tablecounts.tex}

The lack of any significant variation in the 850\,$\mu$m number counts suggests that the environments of SMG are well--sampled when the population is volume--averaged on scales of $\sim$\,0.5--3~sq.\,degree, corresponding to a projected volume on the order of 0.15\,Gpc$^{3}$ (assuming the majority of SMGs lie in the range $z$\,=\,1.5--6). We stress that subsets of SMGs may still reside in large--scale structures with a correspondingly narrow redshift interval, but that these do not result in significant variation in the counts when integrated over the redshift range probed by an 850\,$\mu$m selection ($z$\,$\lsim$\,6). If such structures do exist within the S2COSMOS source catalogue then they remain interesting in the context of galaxy evolution (e.g.\ \citealt{Smail14,Casey15,Ma15,Lewis18,Oteo18,Miller18}) but their identification requires precise 3--D locations for each SMG, which lies beyond the scope of this paper. Pin-pointing the location of each galaxy that contributes to a source in S2COSMOS catalogue can be achieved with high-resolution interferometric imaging at sub-mm wavelengths (e.g.\ \citealt{Hodge13,Simpson15, Stach18}) and indeed such observations are under-way for the brightest sources in the S2COSMOS source catalogue (Simpson et al.\ in prep). In the meantime we are exploiting machine--learning algorithms applied to multiwavelength data (An et al.\ 2018) to derive a catalog of probable counterparts for further study (An et al.\ in prep).

\subsubsection{Comparison to galaxy formation models}\
\label{subsubsec:galformmod}
Finally, we compare our results to both a phenomenological model and a semi-analytic galaxy formation model. The 850\,$\mu$m number counts, including the effect of blending in SCUBA--2\,/\,JCMT observations \citep{Cowley15}, from the GALFORM semi-analytic model of galaxy--formation \citep{Lacey16} are shown in Figure~\ref{fig:counts}. GALFORM attempts to provide a unified model of galaxy formation that reproduces observational results across a wide range of redshift. In GALFORM the SMG phase predominantly arises due to triggered instabilities in gas--rich discs and the current version of the model \citep{Lacey16} adopts a stellar Initial Mass Function (IMF) in starbursts that while top--heavy ($x$\,=\,1) is close to Salpeter ($x$\,=\,1.35; \citealt{Salpeter55}). The predicted number counts from the GALFORM model show broad agreement with S2COSMOS at the very brightest flux densities, S$_{850}$\,$>$\,7\,mJy (see Figure~\ref{fig:counts}). At fainter flux densities GALFORM over-predicts the observed number counts in the S2COSMOS field by a factor of $\sim$\,1.4--1.6\,$\times$. 

The phenomenological modeling presented by \citet{Bethermin17} represents a fundamentally different approach to modeling galaxy formation and evolution, relative to the physics--based semi-analytic method. Briefly, the \citet{Bethermin17} model combines simple empirical relations estimated from observations of galaxies (e.g.\ stellar mass functions); abundance matching techniques to simulations of dark matter halos; and models of galaxy spectral energy distributions to predict the far--infrared emission for galaxies (not including the effect of blending in the map). By its nature this phenomenological approach has much lower predictive power than a semi-analytic model, but does provide an environment in which to explore biases in observational results. To investigate the accuracy of the \citet{Bethermin17} model we create a simulated SCUBA--2 image based on the output of the model and compare this to the S2COSMOS survey. First, we create a simulated image at 850\,$\mu$m using the position and brightness of the predicted sources. Next, the simulated image is convolved with the empirical SCUBA--2 PSF and realistic noise is included by co-adding the resulting map with a randomly selected S2COSMOS jackknife image. Finally, we analyze the simulated SCUBA--2 map in an identical manner to the S2COSMOS survey: sources were extracted at $>$\,4\,$\sigma$ and the resulting catalogue used to create the simulated number counts after applying completeness and deboosting corrections. Overall, the counts extracted from the \citet{Bethermin17} empirical--based model appears to be in close agreement with the single--dish 850\,$\mu$m number counts at $S_{850}$\,$>$\,2\,mJy (see Figure~\ref{fig:counts}), suggesting that it does not need to be recalibrated on the basis of the counts derived here.

\subsection{Environments of S2COSMOS sources}
\label{subsec:stats}
S2COSMOS has identified 1020 sub-mm sources across our {\sc main} COSMOS survey and provides a statistically--robust sample with which to characterize the SMG population. We currently lack complete interferometric imaging of the S2COSMOS sources, so instead we now exploit the optical--to--near-infrared imaging of the field to search for galaxies around the S2COSMOS source positions, representing statistical associations of galaxies with the SMGs.

To determine if there is an excess of a particular type of galaxy around the S2COSMOS positions we use the catalog of optical--\,/\,near-infrared--selected galaxies in the COSMOS field presented by Laigle et al. (2016; COSMOS15). Briefly, \citet{Laigle16} present multi--band photometry for all sources that are detected in an ultra-deep $zJYHK_{s}$ stacked image of the field. The depth of the stacked image varies across the field, primarily due to changes in sensitivity across the deep and ultra-deep regions of the UltraVISTA imaging, leading to variations in the surface density of detected sources. However, the $K_{s}$--band number counts constructed from the UltraVISTA deep and ultra-deep regions are consistent at $K_{s}$\,$\le$\,24.5 \citep{Laigle16} and as such we adopt this selection limit throughout our analysis. In addition, we retain any sources with [3.6\,$\mu$m]\,$\le$\,25.0\,mag (SNR\,$\gsim$\,5) noting that this limit is chosen to improve the completeness level of the catalog for massive ($\gsim$\,10$^{10}$\,$\Msol$) systems located towards higher redshift (see \citealt{Bourne17,Davidzon17}). At these limits we estimate that the source catalog is $\gsim$\,90\,$\%$ complete for (low--obscuration) galaxies with stellar masses $\ge$\,10$^{10}$\,$M_{\odot}$ over $z$\,=\,0--2.5, and $\ge$\,3\,$\times$\,10$^{10}$\,$M_{\odot}$ to $z$\,=\,4 \citep{Laigle16}\,\footnote{We verified our estimate for the mass completeness of the COSMOS15 catalog using an empirical comparison to catalogs extracted from the $JH$--selected 3D--${\it{HST}}$ \citep{Momcheva16} and the $K$--selected ZFOURGE \citep{Straatman16} surveys: the 3D--$\it{HST}$ and ZFOURGE imaging provides coverage over 140--180 sq.\ arcmin within the COSMOS field at a 5\,$\sigma$ limiting depth of 26.1 ($H$) and 25.5\,mag ($K$) and are expected to be mass--complete to over our range of interest in redshift ($z$\,=\,0--4)}. We stress that the estimated completeness levels are sensitive to source reddening, with the most obscured sources often undetected in optically--selected catalogues (e.g.\ \citealt{Chen14, Simpson14}). Thus, while we adopt these redshift and stellar mass bounds we caution that the completeness should be considered an upper limit. Physical properties (e.g.\ photometric redshift, stellar mass) are provided by \citet{Laigle16} for each source, and are derived from modeling the available 30--band photometry spanning near--ultraviolet to IRAC\,/\,8.0\,$\mu$m wavelengths. 

\begin{figure*}
  \includegraphics[width=0.48\textwidth]{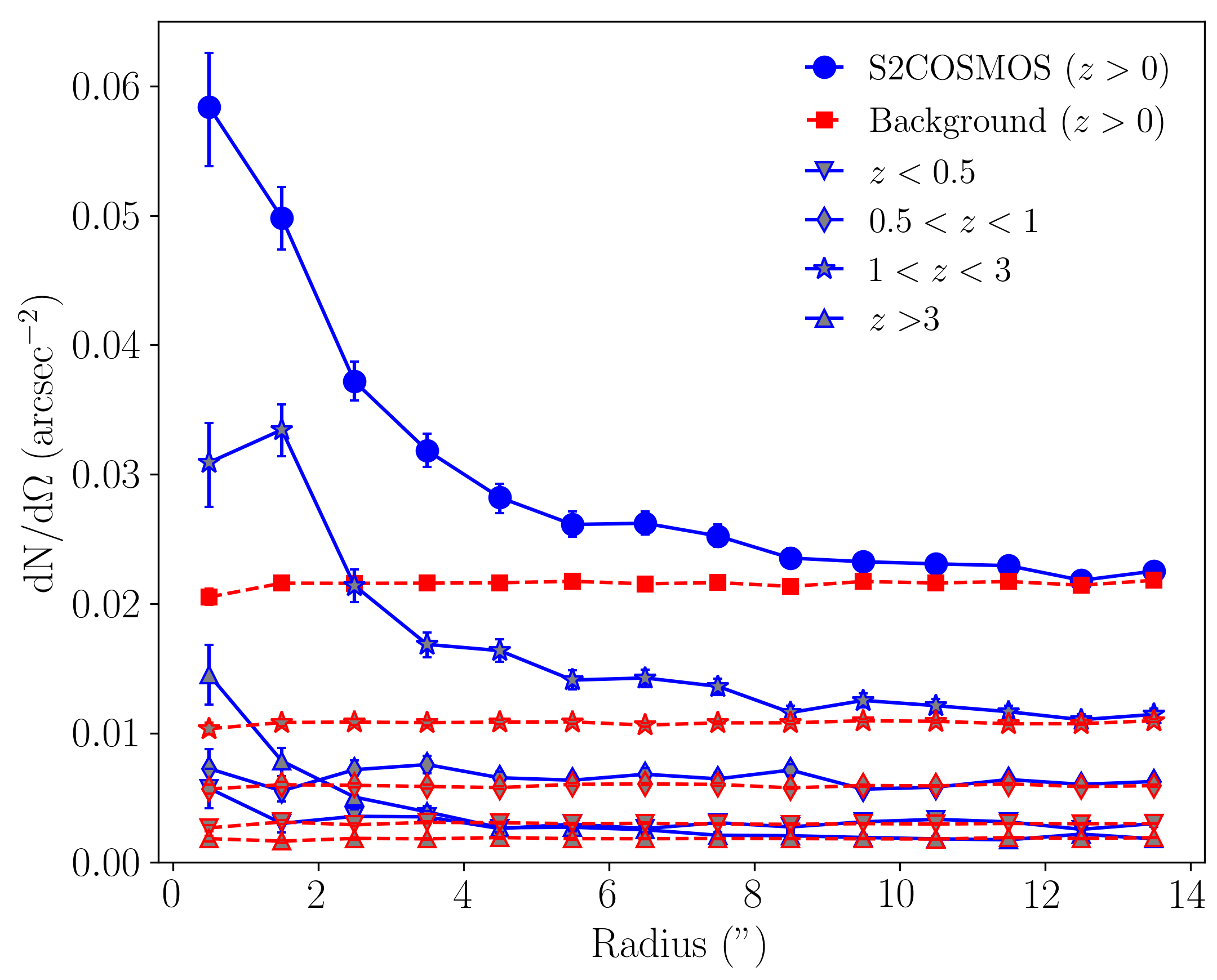}
  \hfill
  \includegraphics[width=0.48\textwidth]{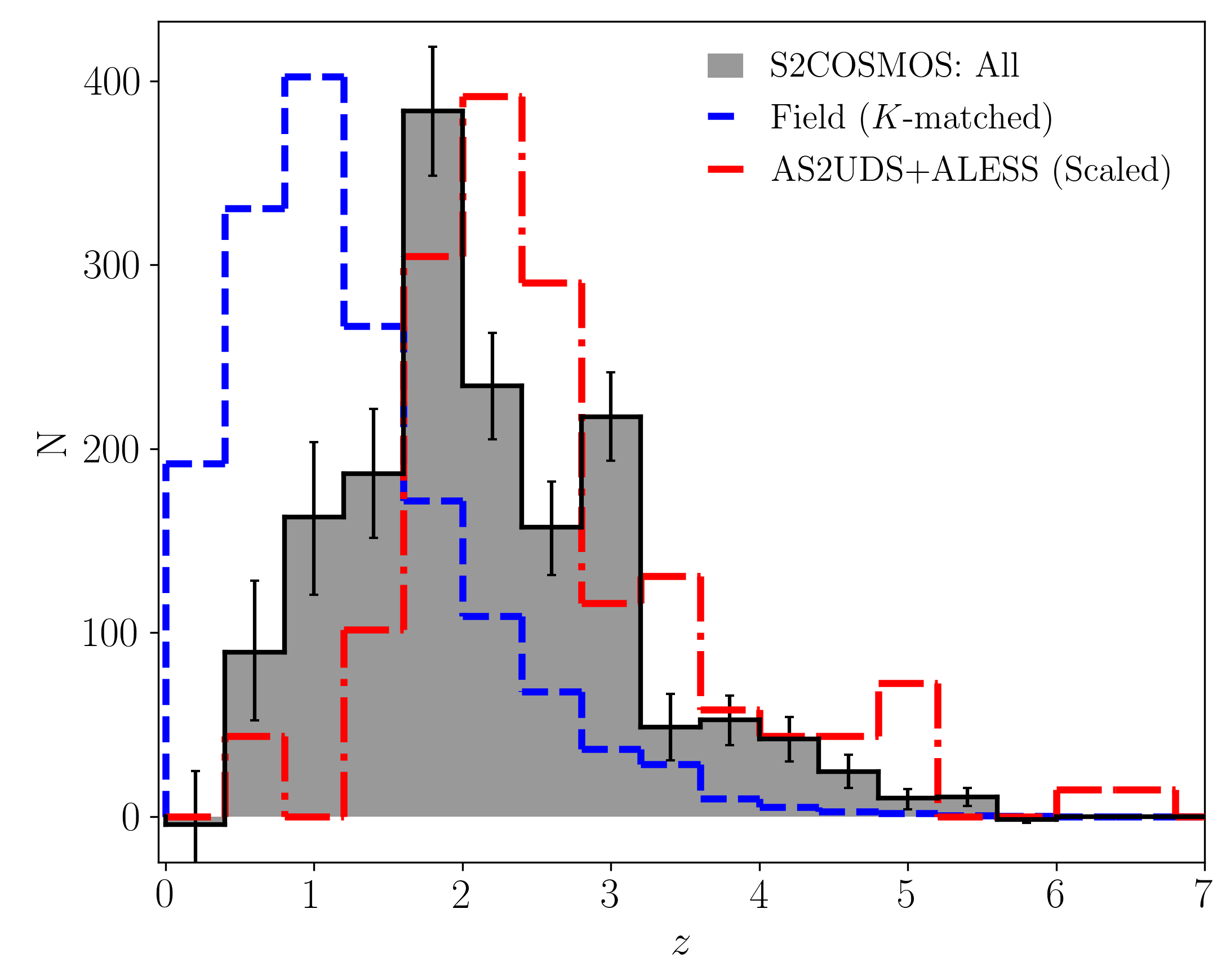}
  \caption{ {\textit{Left:}} The radially--averaged surface density of near--infrared--selected galaxies around S2COSMOS source positions, and separated by photometric redshift. The expected background level of field galaxies for each redshift range is shown (red symbols) and is constructed by considering the distribution of sources around 15000 random positions in the field. The radially--averaged surface density distribution around S2COSMOS positions shows a clear excess above the background level, representing near--infrared--selected SMGs and\,/\,or companion sources. The radial density profile of the galaxies associated with S2COSMOS sources is centered on the SCUBA--2 positions and declines out to a radius of $\sim$\,13$''$ corresponding to $\sim$\,100\,kpc in projected distance. Integrating the measured ``excess'' at $R$\,$<$\,13$''$ we calculate that, on average, there are 2.0\,$\pm$\,0.2 near--infrared--selected galaxies ($K_{s}$\,$\le$\,24.5 or [3.6\,$\mu$m]\,$\le$\,25.0\,mag) associated with each S2COSMOS source, with 73\,$\pm$\,3$\pc$ located at 1\,$<$\,$z$\,$<$\,3. {\textit{Right:}} The photometric redshift distribution for the measured galaxy excess around S2COSMOS positions, representing near--infrared--selected SMGs and\,/\,or companions. For comparison, we show the redshift distribution for a $K$--band--magnitude matched sample of field galaxies and 112 near--infrared-detected SMGs that were identified in ALMA imaging of single-dish--identified sub-mm sources (\citealt{Simpson14,Simpson17}). The S2COSMOS ``excess'' galaxies lie at a median redshift of $z$\,=\,2.0\,$\pm$\,0.1, placing these sources at significantly higher redshift than field galaxies of comparable $K$--band magnitude ($z$\,=\,1.1\,$\pm$\,0.1) and in reasonable agreement with near--infrared detected samples of ALMA--identified SMGs (median $z$\,=\,2.3\,$\pm$\,0.1; \citealt{Simpson14,Simpson17}), confirming that our S2COSMOS survey has located a population of starburst galaxies (SMGs) at high redshift ($z$\,$\gsim$\,1).}
\label{fig:dndgal}
\end{figure*}

As SMGs typically have extremely red colors at optical--to--near--infrared wavelengths (e.g.\ \citealt{Smail99,Ivison01,Frayer04,Yun08,Hainline11,Michalowski12b,Simpson14}) we have limited our analysis to the 998 S2COSMOS {\sc{main}} sources that lie within the UltraVISTA\,/\,$K_{s}$ footprint of the COSMOS field (Figure~1; \citealt{McCracken12}). We remove a further 164 S2COSMOS sources that lie within regions that were masked during the construction of the COSMOS15 catalogue, leaving a sample of 834 {\sc{main}} sub--mm sources for analysis. These sources have a median deboosted flux density of $S_{850}$\,=\,4.0\,$\pm$\,0.1\,mJy and are representative of the overall S2COSMOS {\sc main} sample ($S_{850}$\,=\,4.1\,$\pm$\,0.1\,mJy). We measure the average surface--density of galaxies around each S2COSMOS position and show this as a function of angular offset in Figure~\ref{fig:dndgal}. The surface density of near--infrared--selected galaxies peaks at the location of the S2COSMOS sources and steadily declines with increasing distance from each source before flattening and approaching the background level at $R$\,$\sim$\,13$''$ ($\sim$\,100\,kpc at $z$\,$\sim$\,2). The peak in the measured surface-density distribution of galaxies around the S2COSMOS positions confirms that at least some of the galaxies associated with the sub-mm sources are detectable in the COSMOS15 catalog. However, our measurement contains a  ``background'' contribution due to field galaxies that lie along the line-of-sight to each S2COSMOS source. To estimate the galaxy background level we construct a sample of 15000 randomly--selected positions that are located within $R$\,=\,30--60$''$ of S2COSMOS sources. Using a local estimate for the background measurement ensures that we account for any correlation between large-scale structure (e.g.\ \citealt{Scoville13,Darvish17}) and the spatial variation in instrumental sensitivity across the S2COSMOS survey. The average surface density of galaxies around these random positions is measured and is taken to represent the galaxy background level around the S2COSMOS sources (Figure~\ref{fig:dndgal}). 

As shown in Figure~\ref{fig:dndgal}, we find a significant excess of galaxies around the position of S2COSMOS sources that declines steadily with increasing distance from the sub--mm emission, until reaching the background level at a radius of $\sim$\,13$"$ ($\sim$\,100\,kpc projected). Indeed, integrating the surface density of galaxies around the S2COSMOS positions we determine that there is an average excess of 2.0\,$\pm$\,0.2 galaxies ($K_{s}$\,$\le$\,24.5 or [3.6\,$\mu$m]\,$\le$\,25.0\,mag) within a 13$''$ radius of each S2COSMOS source, after accounting for the background contribution (Figure~\ref{fig:dndgal}). The measured excess is marginally higher than that reported by \citet{Smith17}, who determine an excess of 1.5\,$\pm$\,0.1 $K_{s}$\,$\le$\,24.6 sources within 12$''$ of S2CLS sources in the UKIDSS UDS. To understand the properties of these excess galaxies we estimate their redshift distribution by constructing the full distribution for all galaxies within a 13$''$ radius of each S2COSMOS source and subtracting the expected background contribution, as determined from our analysis of randomly--selected positions. We find that 72\,$\pm$\,3\,$\pc$ and 16\,$\pm$\,2\,$\pc$ of these excess galaxies lie at a redshift of 1\,$<$\,$z$\,$<$\,3 and $z$\,$>$\,3, respectively, significantly higher than the 49\,$\pm$\,1\,$\pc$ and 4.0\,$\pm$\,0.2\,$\pc$ estimated for a $K$--band magnitude matched sample. Note that we correct the redshift estimates for the background population by binning the SMG and background populations into $\Delta$\,$z$\,=\,0.25 bins, and subtracting the average distribution 

The galaxy excess around the S2COSMOS positions has a median photometric redshift of $z$\,=\,2.0\,$\pm$\,0.1, in reasonable agreement with the median of $z$\,=\,2.3\,$\pm$\,0.1 determined for near--infrared detected samples of ALMA--identified SMGs (e.g.\ \citealt{Simpson14,Simpson17}). However, we stress that our analysis is sensitive to both SMGs and any other associated sources, either at the same redshift or along the line--of--sight, and that the lower median redshift determined here may reflect an increasing sensitivity towards fainter companions at lower redshifts or the subtle effect of weak lensing.

\begin{figure*}
\includegraphics[width=0.48\textwidth]{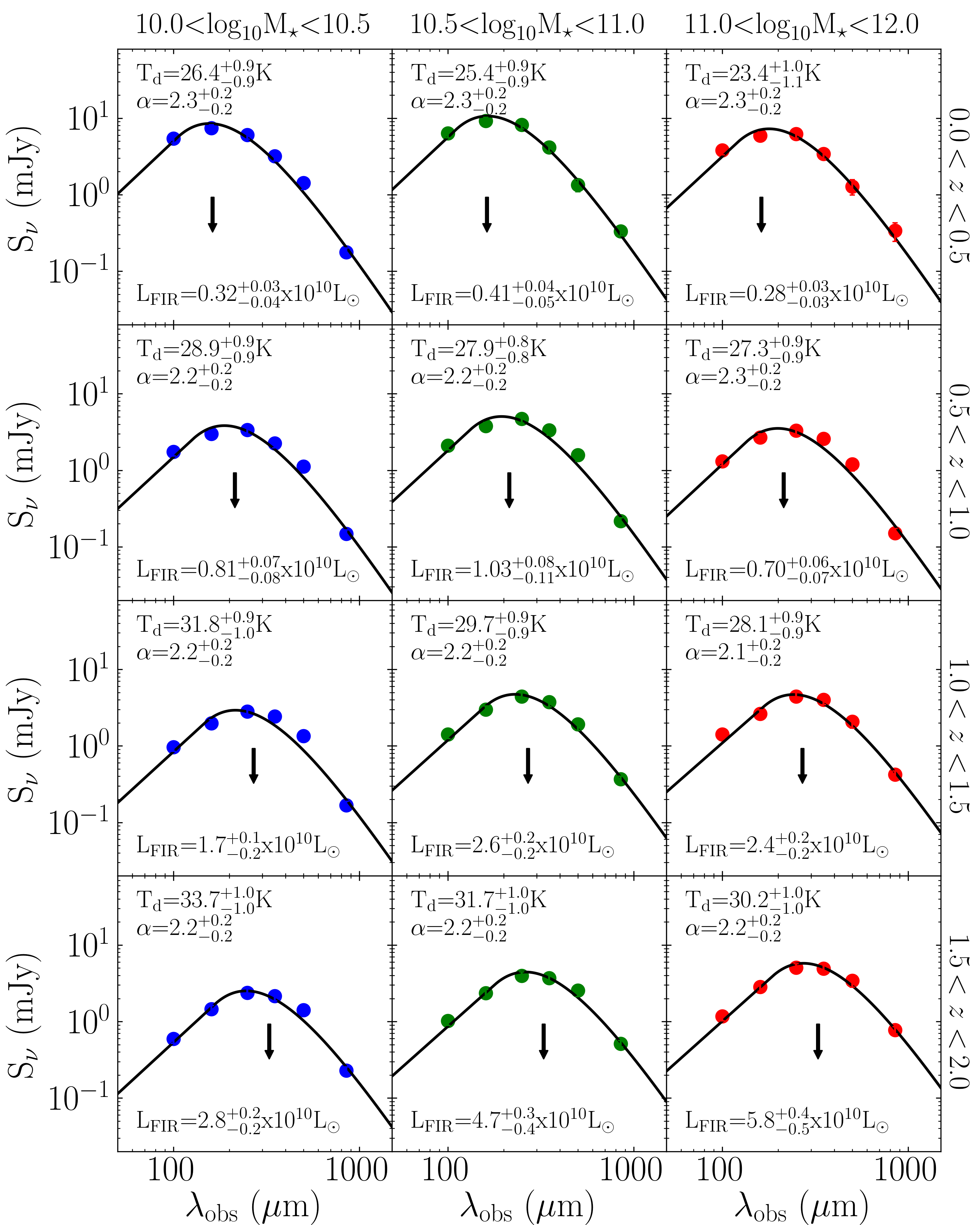}
\hfill
\includegraphics[width=0.48\textwidth]{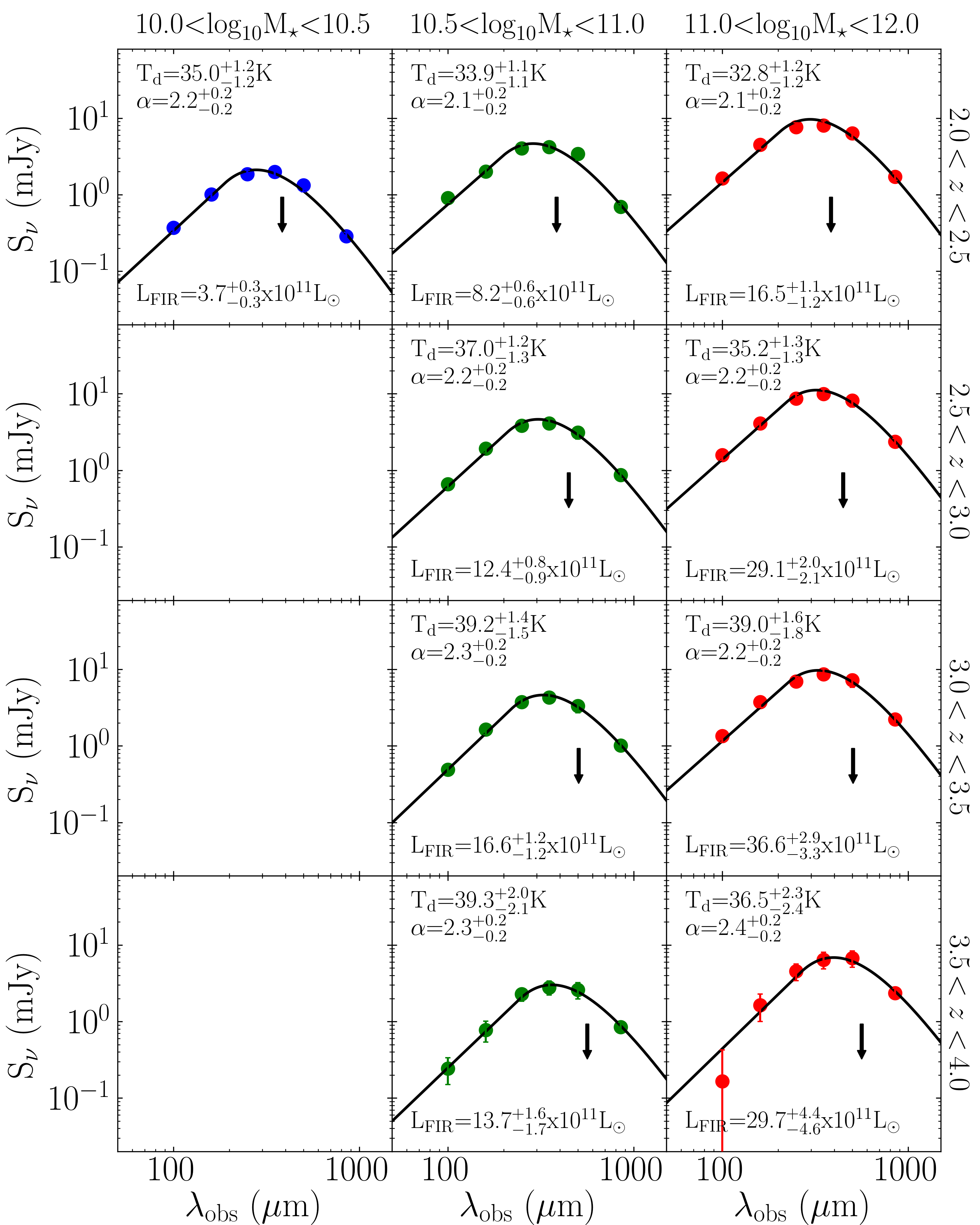}
  \caption{The average stacked flux density at 100--850\,$\mu$m of mass--selected galaxies in the COSMOS field in three subsets of stellar mass and eight subsets in redshift. A solid line in each panel represents the best-fit model, comprising an optically--thin, modified blackbody (dot--dash) and power--law function at mid--infrared wavelengths, fit to the observed photometry. The average far--infrared SED of mass--selected galaxies evolves strongly with redshift, with the rest-frame peak of the dust emission decreasing steadily from 120\,$\pm$\,5$\mu$m to 80\,$\pm$\,5\,$\mu$m between $z$\,=\,0.25--3.75; an arrow on each panel represents the observed wavelength corresponding to average rest-frame peak wavelength of the lowest redshift subset. The typical evolution out to $z$\,=\,4 in peak wavelength corresponds to an increase of 13\,$\pm$\,2K in the luminosity--weighted dust temperature, under the assumption of optically--thin emission, and indicates that the nature of star-forming regions within galaxies at fixed stellar mass evolves strongly with lookback time.}
\label{fig:sedfits}
\end{figure*}

To search for trends in the redshift distribution of S2COSMOS associated galaxies with 850\,$\mu$m flux density we split the S2COSMOS sample into subsets at $S_{850}$\,$=$\,2--4, 4--6, 6--8 and $>$8\,mJy and repeat our analysis. We identify an excess of galaxies in each flux bin and a weak dependence between the median redshift and flux density of each sample: for sources at $S_{850}$\,=\,2--4\,mJy we estimate a median redshift of $z$\,=\,1.9\,$\pm$\,0.1, increasing slightly to $z$\,=\,2.0\,$\pm$\,0.1, 2.2\,$\pm$\,0.2, and 2.4\,$\pm$\,0.2 at $S_{850}$\,$=$\,4--6\,mJy, $=$\,6--8\,mJy, and $>$\,8\,mJy, respectively. While this hints that more luminous 850\,$\mu$m sources lie at higher redshifts (e.g.\ \citealt{Stach19}), we caution that a simple explanation for our results is that more intense starbursts may be intrinsically brighter at optical--to--near--infrared wavelengths and, as such, can be traced to higher redshift at a fixed observed luminosity (e.g.\ \citealt{Simpson14}).

Finally, we consider whether the S2COSMOS sample is strongly contaminated by sources that are gravitationally--lensed by foreground galaxies. If strong gravitational lensing affects a significant fraction of our sample then we can expect to see an excess of foreground galaxies in the vicinity of the S2COSMOS positions. However, our analysis shows no evidence for a strong excess of galaxies with photometric redshifts in the range $z_{\mathrm{phot}}$\,$<$\,0.5 around either the full S2COSMOS sample, or the subset with flux densities of $S_{850}$\,$=$\,6--8 or $>$\,8\,mJy. The absence of a correlation with the foreground population is consistent with our analysis of the S2COSMOS number counts and indicates that strong--lensing by low redshift sources ($z$\,$\lsim$\,0.5) is not a major concern for the majority of our sample. We do measure a significant excess of galaxies at 0.5\,$<$\,$z_{\mathrm{phot}}$$<$\,1 around S2COSMOS positions but disentangling any possible gravitationally--lensed S2COSMOS sources from sub-mm sources that truly lie at these redshifts is challenging. However, we comment that the radial distriubtion of the galaxy excess is more uniform for sources that lie at 0.5\,$<$\,$z_{\mathrm{phot}}$$<$\,1, relative to $z_{\mathrm{phot}}$\,$>$\,1, indicating a larger angular separation between the galaxy excess at 0.5\,$<$\,$z_{\mathrm{phot}}$$<$\,1 and the S2COSMOS sources. While this increase in the average separation between the S2COSMOS sources and associated near--infrared galaxies at 0.5\,$<$\,$z_{\mathrm{phot}}$$<$\,1 may be a potential indicator of gravitational lensing we again stress that it may also reflect a higher sensitivity to near--infrared--selected companions at lower redshift. Thus, we caution that we cannot rule out the presence of weak--lensing by low--redshift sources\,/\,foreground structures (e.g.\ \citealt{Almaini05, Aretxaga11, Bourne14}), or strong--lensing systems at $z$\,$\gsim$\,1 (see \citealt{Vieira10}), and this will be investigated in further detail in future work (An et al in prep.; Simpson et al in prep.).

\subsection{Properties of mass--selected galaxies}
\label{subsec:simstack}
We have demonstrated that the S2COSMOS survey provides measurements of the properties of luminous strongly star-forming galaxies over a wide range of cosmic history. However, while these 850\,$\mu$m--luminous sources are a key population at high redshift, the majority of the galaxy population lies below the detection threshold of our imaging. Thus, in the following we use a stacking analysis to extend our analysis and estimate the average far--infrared properties of mass--selected sources. 

To construct a sample for a stacking analysis we again use the catalogue of optical--to--near-infrared selected galaxies presented by \citet{Laigle16}, enforcing the same selection limits described in \S\,\ref{subsec:stats}. To search for trends in the far-infrared properties of these sources as a function of their stellar mass and redshift we split our sample into three bins at log$_{10}$\,$M_{\star}$\,=\,10.0--10.5,\,10.5--11.0,\,11.0--12.0 and eight $\Delta$\,$z$\,=\,0.5 bins from $z$\,=\,0--4 ($N_{\mathrm{bin}}$\,=\,100--8100; median 1600). Given the redshift range of our study we do not attempt to split our sample into ``star--forming'' and ``quiescent'' systems based on their optical--to--near--infrared colors, in contrast to many previous studies (e.g.\ \citealt{Magdis12,Viero13,Santini14,Bethermin15,Schreiber15}). These color cuts are known to mis-classify dust--obscured star-forming systems as quiescent (\citealt{Smail02,Toft05,Dunlop07,Caputi12,Simpson17,Eales18}), with the failure rate estimated at $\sim$\,25--50\,$\pc$ by $z$\,$\gsim$\,3 \citep{Chen16b,Schreiber18b}. 

At the coarse resolution achieved in 850\,$\mu$m observations with the JCMT, source blending within the beam is a major source of bias in any stacking analysis. To address this we determine the stacked 850\,$\mu$m flux density for each subset using {\sc simstack} \citep{Viero13}, a publicly--available code that attempts to correct for the clustering of sources within the {\sc fwhm}\,=\,15$''$ scale of the JCMT beam. Briefly, {\sc simstack} models each ``subset'' of an input catalogue as a single ``layer'', regressing each ``layer'' simultaneously with the true sky map to estimate the average flux density for each subset (see \citealt{Viero13}). To improve the fitting process we modify the {\sc simstack} code to also simultaneously model the background level on each image and to account for regions in the optical\,/\,near--infrared images that were masked in the construction of the COSMOS15 catalogue. These changes are verified in the following section using a suite of simulated maps that match the area coverage and masking strategy of the COSMOS2015 catalogue. Finally, the associated uncertainty on each stacked flux density is determined by combining the measurement uncertainty, the uncertainty determined from a bootstrap analysis, and the expected uncertainty on the flux calibration (8\,$\pc$; \citealt{Dempsey13}).

Using our updated version of the {\sc simstack} code we identify 850\,$\mu$m emission from all galaxy subsets that are considered in our stacking analysis at a SNR\,=\,4--30 (median SNR\,=\,14, and not including systematic flux calibration uncertainty). To construct the global far--infrared properties of our sample we extend our stacking analysis to the available {\it Herschel}\,/\,PACS and {\it Herschel}\,/\,SPIRE imaging of the COSMOS field that was obtained as part of the PACS Evolutionary Probe (PEP; \citealt{Lutz11}) and the HerMES \citep{Oliver12} surveys, respectively. The PACS imaging at 100 and 160\,$\mu$m achieves a typical $1$--$\sigma$ instrumental sensitivity of 2--4\,mJy\,beam$^{-1}$ (FWHM\,=\,7--11$''$), while the SPIRE 250, 350 and 500\,$\mu$m maps reach a median $1$--$\sigma$ instrumental sensitivity of 1.7--2.0\,mJy\,beam$^{-1}$ (FWHM\,=\,18--35$''$). The relative astrometry of each image is confirmed by stacking on the map at the position of 3\,GHz\,/\,VLA sources (\citealt{Smolcic17}; \S~\ref{subsubsec:ast}). We estimate the average 100–-500\,$\mu$m emission from each subset, and its associated uncertainty, following the same stacking procedure that was employed at 850\,$\mu$m, assuming a flux calibration uncertainty of 5.0$\pc$\footnote{http://herschel.esac.esa.int/Docs/PACS/html/pacs$\textunderscore$om.html} and 5.5$\pc$\footnote{http://herschel.esac.esa.int/Docs/SPIRE/html/spire$\textunderscore$om.html} for PACS and SPIRE imaging, respectively.

Our stacking analysis identifies strong emission at 100--850\,$\mu$m from each mass-selected subset at a median detection significance of 20\,$\sigma$ (see Figure~\ref{fig:sedfits}). To verify the accuracy of our stacking results we repeat our analysis on simulated maps constructed from the phenomenological model of galaxy formation presented by Bethermin et al.\ (2017; see \S\,\ref{subsubsec:galformmod}). We identify and correct for systematic offsets of 3--11\,$\pc$ at 100--850\,$\mu$m, which we attribute to residual blending issues with galaxies at lower stellar masses, and incorporate scatter in these corrections into the associated uncertainties on each of our stacked flux density measurements. Note that if the average flux density of each subset is taken as weighted mean at the position of each source then the correction factors for blending are 3--40\,$\times$ higher, confirming the strength of the simultaneous stacking approach employed in the {\sc simstack} routine.

To characterize the far--infrared emission from each stacked subset we initially model the observed photometry with a single--temperature, optically--thin, modified blackbody (mBB) function

\begin{equation}
S_{\nu_{\mathrm{obs}}} \propto  \nu_{\mathrm{rest}}^{\beta}B \left( \nu_{\mathrm{rest}},T_{\mathrm{d}} \right),
\end{equation}

\noindent where $B\left(\nu_{\mathrm{rest}},T_{\mathrm{d}}\right)$ represents the Planck function and $\beta$ the dust emissivity which we assume to be 1.8 (\citealt{Planck11}). A single dust--temperature, modified blackbody is known to under-predict the short--wavelength dust emission from infrared--bright sources \citep{Blain03} and this is evident in our stacked SEDs (Figure~\ref{fig:sedfits}). As such, we adopt a power--law SED model (S$_{\nu}$\,$\propto$\,$\nu^{-\alpha}$) in the mid--infrared following the prescription of \citet{Blain03}. Thus, our SED model contains three parameters ($T_{\mathrm{d}}$, $\alpha$, and a normalization, $N$) and their best--fit values and associated uncertainties to the observed photometry from each subset are determined using a Monte Carlo Markov Chain sampler ({\sc emcee}; \citealt{EMCEE13}) following the procedure presented in \citet{Simpson17}. To ensure that the SED fitting returns physically--motivated results we place a Gaussian prior on $\alpha$ at $\alpha$\,=\,2.3\,$\pm$\,0.2, which is motivated by modeling the observed 100--850\,$\mu$m photometry of spectroscopically--confirmed, high--redshift SMGs (zLESS; \citealt{Danielson17}) and is consistent with previous studies (e.g.\ \citealt{Blain03,Casey13}). 

\subsubsection{Redshift Evolution in the average SED}
\label{subsec:tdevo}
In Figure~\ref{fig:sedfits} we present the stacked photometry and best--fit SED model for each of our galaxy subsets. From these stacks we identify two clear redshift trends in the far--emission from mass--selected sources. First, the relative luminosity between each stellar--mass subset evolves strongly with redshift; at $z$\,$\lsim$\,1.5, lower mass galaxies (log$_{10}$\,$M_{\star}$\,$<$\,11.0) are on average more luminous at far--infrared wavelengths than the most massive systems, but this is reversed by $z$\,$\gsim$\,1.5. This trend of `downsizing' with redshift of the luminosity and, by proxy, star-formation rate (SFR), of mass--selected galaxies is well--known (e.g.\ \citealt{Cowie96,Thomas10,Sobral11,Karim11}) and may reflect redshift evolution in the fraction of ``active'' and ``passive'' galaxies at a fixed stellar mass (e.g.\ \citealt{Mortlock15,Davidzon17}). Second, the rest--frame peak of the dust SED shifts, on average, from 120\,$\pm$\,5$\mu$m to 80\,$\pm$\,5$\mu$m between $z$\,=\,0.25--3.75, corresponding to an increase of 13\,$\pm$\,2\,K in luminosity--weighted dust temperature for a single-temperature, optically--thin mBB ($T_{\mathrm{d}}$\,=\,25\,$\pm$\,1\,K to 38\,$\pm$\,2\,K; see Figure~\ref{fig:sedfits}).

An increase in the rest--frame peak wavelength of the dust SED with redshift is in broad agreement with observations of both individual sources (e.g.\ \citealt{Hwang10,Swinbank13,Strandet16,Fudamoto17,Cooke18}) and prior stacking analyses of mass-- and SFR--selected galaxies (\citealt{Magdis12,Viero13,Magnelli14,Bethermin15,Schreiber18}). Indeed, \citet{Schreiber18} present a stacking analysis of the far--infrared emission from optically--selected ``star-forming'' sources (log$_{10}$\,$M_{\star}$\,$>$\,9.5) that were identified across the 0.2 sq.\ degree CANDELS fields, finding that the rest-frame peak of the emission shifts from 100\,$\mu$m to 65\,$\mu$m from $z$\,=\,0.25--3.75. The overall trend of increasing dust temperatures for each of our subsets is in broad agreement with that presented by \citet{Schreiber18}, albeit with a systematic offset towards longer wavelengths that likely results from differences in selection (see \S\,\ref{subsubsec:lfirmod}) and SED fitting technique, and confirms the apparent increase in the peak wavelength of the dust SED of mass--selected galaxies with redshift (see also \citealt{Viero13,Magnelli14,Bethermin15}). 

\begin{figure}
\includegraphics[width=0.49\textwidth]{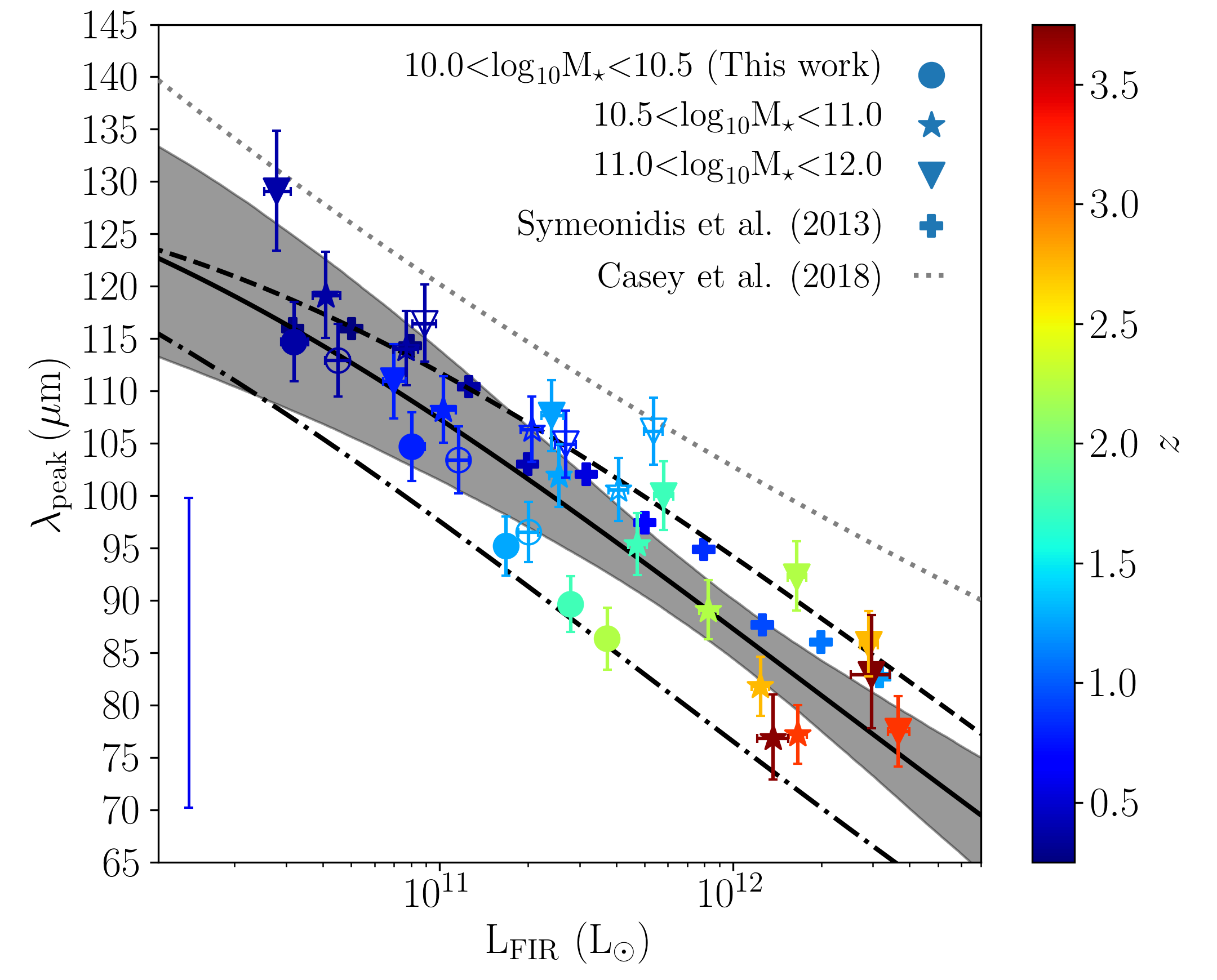}
  \caption{The relation between far--infrared luminosity and rest-frame peak wavelength of the dust SED for mass--selected galaxies in the COSMOS field, in subsets of stellar mass and redshift. Open symbols represent our stacking results for ``active'' galaxies in the COSMOS field at $z$\,$<$\,1.5 that were selected based on the $NUV$--$r$ and $r$--$J$ colors. For comparison we show a sample of infrared--bright sources at $z$\,$=$\,0--1 (\citealt{Symeonidis13}; typical uncertainty is shown in the lower left) and the best-fit relation to a heterogeneous sample of far--infrared--selected systems at $z$\,=\,0--6 \citep{Casey18}. We construct a simple model for the expected trend between $\lambda_{\mathrm{p}}$ and L$_{\mathrm{FIR}}$, which shown for galaxies at log$_{10}$\,$M_{\star}$\,$=$\,10.0--10.5 (dot-dash), 10.5--11.0 (solid), and 11.0--12.0 (dashed). Note tracks are normalized to match our stacking result for galaxies at log$_{10}$\,$M_{\star}$\,$=$\,10.5--11.0 and $z$\,=\,2.0--2.5. This simple model provides a reasonable representation of the relation between far--infrared luminosity and rest-frame peak wavelength of our stacked subsets, including the dependence on stellar mass, although we caution that there is a large uncertainty associated with each track (shaded region; log$_{10}$\,$M_{\star}$\,$=$\,10.5--11.0).}
\label{fig:td_survey}
\end{figure}

In Figure\,\ref{fig:td_survey} we investigate the relation between the rest--frame peak of the dust SED and far--infrared luminosity (8--1000\,$\mu$m) of each of our stacked subsets. We find that for our sample of mass--selected sources the rest-frame peak wavelength of the dust SED decreases with increasing luminosity (and redshift), and that there is a broad decrease in the rest-frame peak wavelength with stellar mass, at a fixed far--infrared luminosity. A relation between far--infrared luminosity and dust temperature (L$_{\mathrm{FIR}}$--T$_{\mathrm{dust}}$), or peak wavelength (L$_{\mathrm{FIR}}$--$\lambda_{\mathrm{peak}}$) was first identified in observations of infrared--luminous sources in the local Universe (e.g.\ \citealt{Dunne00,Chapman03}) and can be interpreted as evolution in the physical properties of star--forming systems as a function their infrared luminosity, or star-formation rate. Comparing to sources at lower redshift, we find that the the L$_{\mathrm{FIR}}$--$\lambda_{\mathrm{peak}}$ relation for our stacked subsets is in good agreement with that determined by \citet{Symeonidis13} for a sample of {\it Herschel} PACS\,/\,SPIRE--selected infrared--luminous sources at $z$\,$=$\,0--1, although we caution that we have not attempted to match the low redshift sample in stellar mass. Thus, to first--order redshift evolution in the peak wavelength for mass-selected sources is consistent with the well--established L$_{\mathrm{FIR}}$--$\lambda_{\mathrm{peak}}$ relation and redshift evolution in average far--infrared luminosity of galaxies selected at a fixed stellar mass (see Figure\,\ref{fig:sedfits}), and should not be interpreted in terms of a global $\lambda_{\mathrm{peak}}$(T$_{\mathrm{dust}}$)--$z$ relation.

\citet{Casey18} recently presented a best--fit L$_{\mathrm{FIR}}$--$\lambda_{\mathrm{peak}}$ relation for a heterogeneous sample of far--infrared--to--millimeter--selected galaxies at $z$\,=\,0--6, which was subsequently employed in a phenomenological model of galaxy evolution designed to estimate the number counts and redshift distribution of sub--mm\,/\,mm sources \citep{Casey18b}. The L$_{\mathrm{FIR}}$--$\lambda_{\mathrm{peak}}$ relation presented by \citet{Casey18} lies systematically above our results by $\sim$\,5--10\,$\mu$m, at a fixed far--infrared luminosity, with the offset increasing to $\sim$\,15--20\,$\mu$m for the most luminous systems (see Figure\,\ref{fig:td_survey}; equivalent to $\Delta$\,$T_{\mathrm{d}}$\,$\sim$\,3--7\,K). The sample of infrared--selected sources analyzed by \citet{Casey18} has a complex, redshift--dependent, selection function and we suggest that this may be the primary driver for the discrepancy with the results presented here, although note that we discuss the limitations of our stacking analysis in \S\,\ref{subsubsec:lfirmod}. Indeed, at high far--infrared luminosity the sample constructed by \citet{Casey18} is dominated by 1.4\,mm--selected sources, which may introduce a bias towards lower dust temperature (see \citealt{Blain02,Swinbank13}). Investigating this discrepancy further is beyond the scope of this work but we highlight that a systematic reduction in the normalization in the L$_{\mathrm{FIR}}$--$\lambda_{\mathrm{peak}}$ relation presented by \citep{Casey18b} would decrease the estimated surface density of sub--mm\,/\,mm sources predicted by their model, although the magnitude of the change is sensitive to the redshift evolution of the assumed far--infrared luminosity and stellar mass functions, and the relative mapping between the two. 

\subsubsection{A simple model of L$_{\mathrm{FIR}}$--$\lambda_{\mathrm{peak}}$}
\label{subsubsec:lfirmod} 
In Figure\,10 we show that peak wavelength of the stacked emission from mass--selected sources decreases with both far--infrared luminosity (redshift) and stellar mass. The far--infrared emission from our stacked subsets is expected to arise from dust grains that are in equilibrium with their local radiation field. Under the assumption that the dust grains are at a single temperature, and are optically--thin to far--infrared emission, they will thermally radiate at a peak wavelength given by 

\begin{equation}
\lambda_{\mathrm{peak}}^{4+\beta} \propto T_{\mathrm{d}}^{-4-\beta} \propto  M_{\mathrm{dust}} L_{\mathrm{FIR}}^{-1}.
\end{equation}

Thus, the temperature of a source increases if the same mass of dust absorbs a higher intensity of radiation, shifting the peak of the reprocessed emission towards shorter wavelengths, and remains constant if the dust mass scales linearly with far--infrared luminosity. In this simple approximation, our stacking results indicate that the average radiation field per unit dust mass varies across our sample, and increases in systems at higher luminosity (redshift) and lower stellar mass.

The dust mass of a galaxy is expected to evolve in tandem with the global properties of the system (e.g. metallicity; \citealt{RemyRuyer14}) and, to test whether our stacked SEDs are consistent with this expected evolution, we now creating a simple toy model based on empirically--derived relations. We follow a similar approach to previous studies (e.g.\ \citealt{Magdis12,Magnelli14,Bethermin15}), but for clarity detail each of the relevant assumptions again here. First, the far--infrared luminosity of a galaxy can be related to the total gas mass of the system following the integrated Kennicutt--Schmidt (K-S; \citealt{Schmidt59,Kennicutt98a}) relation

\begin{equation}
L_{\mathrm{FIR}} \propto \mathrm{SFR} \propto M_{\mathrm{gas}}^x,
\end{equation}

where we assume $x$\,=\,1.2\,$\pm$\,0.1 (e.g.\ \citealt{Bothwell13,Sargent14}), adopting an associated uncertainty that reflects varying estimates in the literature \citep{Genzel10,Swinbank12a,Tacconi13,Santini14,Genzel15}. 

Next, we relate the integrated gas mass to the dust mass using the empirical calibration presented by \citet{RemyRuyer14}, who determined the best--fit relation between the dust--to--gas ratio ($\delta_{\mathrm{dgr}}$) and metallicity (Z) for a sample 126 galaxies in the local Universe. Following \citet{RemyRuyer14} we assume their best--fit broken power--law model for the dust--to--gas ratio, based on an $\alpha_{\mathrm{co}}$ conversion factor that includes a dependence on metallicity

\begin{equation}
M_{\mathrm{gas}} \propto  M_{\mathrm{dust}} \delta_{\mathrm{dgr}}\left(\mathrm{Z}\right),
\end{equation}

which, combined with Eq.\,6, yields,

\begin{equation}
\lambda_{\mathrm{peak}}^{4+\beta} \propto \delta_{\mathrm{dgr}}\left(\mathrm{Z}\right) L_{\mathrm{FIR}}^{-\left(x+1\right)}. 
\end{equation}

Finally, to compare our toy model to the far--infrared emission from our stacked subsets we require an estimate of their average metallicity. In the local Universe a tight relation exists between the stellar mass, SFR and metallicity of a galaxy, the so called fundamental metallicity relation (FMR; \citealt{Mannucci10}). We adopt the FMR parameterization presented by Mannucci et al.\ (2010; Eq.\,2) to estimate the metallicity of our mass-selected samples, as a function of far--infrared luminosity, and note that while this FMR was calibrated on galaxies at $z$\,$\lsim$\,0.3 recent observations suggest that it may hold to $z$\,$\sim$\,2.3 (\citealt{Cresci18}). \citet{Troncoso14} present tentative evidence that the metallicity of sources at higher redshift ($z$\,$\sim$\,3.4) may lie 0.4\,$\pm$\,0.2\,dex lower than that expected based on the locally--calibrated FMR. As noted by \citet{Troncoso14} their results may not be representative of the wider galaxy population at these redshifts, but we caution that if the metallicity of galaxies at $z$\,$\gsim$\,2.5 are indeed systematically lower than that predicted by the FMR then our toy model will over-estimate their dust mass, and thus over-estimate the peak wavelength of the dust emission.

In Figure\,\ref{fig:td_survey} we show tracks through the L$_{\mathrm{FIR}}$--$\lambda_{\mathrm{peak}}$ plane that are constructed from combining Eq.\,8 with the FMR. The tracks are normalized to our results for galaxies at log$_{10}$\,$M_{\star}$\,$=$\,10.5--11.0 and $z$\,=\,2.0--2.5, but the relative stellar mass dependence is independent of normalization. As can be seen in Figure\,\ref{fig:td_survey}, this simple model provides a broadly representative description of the observed trends in L$_{\mathrm{FIR}}$--$\lambda_{\mathrm{peak}}$, including the dependence on stellar mass (see also \citealt{Magnelli14}). From our stacking analysis we identify potential ``curvature'' in the L$_{\mathrm{FIR}}$--$\lambda_{\mathrm{peak}}$ plane, at fixed stellar mass, and a ``rotation'' in the trend of $\lambda_{\mathrm{peak}}$ with stellar mass, at a fixed redshift, from $\lambda_{\mathrm{peak}}$ being proportional to M$_{\star}$ at $z$\,$\sim$\,0 to independent of stellar mass at $z$\,$\sim$\,4. It is clear from Figure\,\ref{fig:td_survey} that while the observed curvature is partially reproduced by the model the average mass--selected source at log$_{10}$\,$M_{\star}$\,$>$\,10.5 and $z$\,$\gsim$\,3 does lie below the predicted curve, albeit within the associated uncertainty. \citet{Bethermin15} suggest that such an offset could be explained if galaxies depart from the FMR at high redshift ($z$\,$\gsim$\,2.5) and while this may indeed be true for galaxies at high redshift (e.g.\ \citealt{Troncoso14}) the associated uncertainty on our model tracks means that it is not possible to robustly conclude this with the current data.

The observed trends in the L$_{\mathrm{FIR}}$--$\lambda_{\mathrm{peak}}$ plane may also be due to redshift evolution in the proportion of non--star-forming galaxies in the different sub-samples (e.g.\ \citealt{Davidzon17}) if the far--infrared properties of these ``passive'' galaxies differ from the star-forming population. To estimate the bias that passive, or lower star-formation rate, galaxies may have on our results we split our sample into ``star-forming'' and ``passive'' subsets using the restframe $NUV$--$r$ and $r$--$J$ color cuts proposed by \citet{Ilbert10} and repeat our stacking analysis. We stress that while such color cuts are the only available method to separate our sample into ``passive'' and ``active'' subsets they are known to introduce an artificial bimodality into the galaxy population (see \citealt{Eales18}), and yield ``passive'' samples that suffer increasing contamination from star-forming galaxies with redshift (e.g.\ \citealt{Chen16b}). Thus, to limit the contamination from highly star forming systems we apply the $NUVrJ$--selection to sources that lie at $z$\,$<$\,1.5 and, in doing so, restrict our analysis to sources that have well--defined rest--frame colors.

In Figure\,\ref{fig:td_survey} we show the relation between the rest--frame peak of the dust SED and far--infrared luminosity for ``active'' galaxies at $z$\,$<$\,1.5. We find that the average luminosity of the ``active'' sample is indeed systematically higher than the equivalent mass--selected sample, but that the peak wavelength of the emission remains largely unchanged. The result is a steepening of the L$_{\mathrm{FIR}}$--$\lambda_{\mathrm{peak}}$ relation relative to our mass--selected sample, with the ``rotation'' in the trend of $\lambda_{\mathrm{peak}}$ with stellar mass, at a fixed redshift, largely removed, although note that an overall dependance on stellar mass remains. This highlights that the relationship between the physical properties derived from any stacking analysis at far--infrared wavelengths is sensitive to the underlying, as yet unknown, distribution of source properties. Constructing an unbiased sample of star-forming galaxies is not possible using the available data in the COSMOS field, an issue that becomes evident if we consider that the most massive ``passive'' systems at $z$\,$\sim$\,1.25 have an estimated far--infrared luminosity that is comparable to low--mass, ``active'' galaxies at $z$\,$\sim$\,0.25.

Despite the caveats discussed above, we can use our toy model to provide physical insights on our stacking results and the broad trends that are observed in the L$_{\mathrm{FIR}}$--$\lambda_{\mathrm{peak}}$ plane. Firstly we comment that the gradient of the L$_{\mathrm{FIR}}$--$\lambda_{\mathrm{peak}}$ relation from our toy model is sensitive to changes in the efficiency with which the molecular gas is converted into stars, i.e.\ the slope of the integrated K--S relation. If the star-formation efficiency was invariant across our sample (i.e.\ $x$\,=\,1.0) than our toy model would under-predict the peak wavelength of our stacked dust SEDs at L$_{\mathrm{FIR}}$\,$\lsim$\,3\,$\times$\,10$^{11}$\,$\Lsol$. Conversely, a stronger increase in star-formation efficiency with far--infrared luminosity ($x$\,=\,1.4), would bring the model into closer agreement with our observations. Next, we highlight again that our simple model broadly reproduces the observed stellar--mass dependency in the L$_{\mathrm{FIR}}$--$\lambda_{\mathrm{peak}}$ relation (see also \citealt{Magnelli14}). This stellar mass dependency arises solely from the assumed FMR and DGR--metallicity relations, and would not exist in our model if the assumed DGR was independent of metallicity. As such, our results are consistent with the existence of a DGR--metallicity relation out to $z$\,$\sim$\,4, although we caution that the exact slope of the DGR--metallicity relation that is required to reproduce our stacking results is strongly correlated with the parameterization of the FMR. 

Finally, it is important to comment that our toy model is undoubtedly an over--simplification of the underlying physical processes in these systems, with the shape of the dust SED also expected to be sensitive to the geometry of the emitting dust, pressure of the ISM, properties of dust grains, and the hardness of the stellar radiation field (e.g.\ \citealt{Takeuchi05,Groves08,Narayanan18,Liang19}). Similarly, we have neglected to consider possible systematic biases in the stellar mass estimates of our sample (e.g.\ \citealt{Hainline11}) and, as such, we caution against over-interpreting these results.

\section{Conclusions} 
We have presented S2COSMOS, a deep 850\,$\mu$m survey of the COSMOS field with the SCUBA--2 instrument at the JCMT. The S2COSMOS survey combines 640\,hrs of observations, including 223\,hrs undertaken as an EAO Large Program along with archival data, to map 2.6 sq.\ degree centered on the COSMOS field to a depth of 0.5--3.0\,\mjpb, with a uniform coverage of 1.2\,\mjpb over 1.6\,sq.~degree. The main results from our study are summarized below:

\begin{itemize}

\item We define a {\sc main} survey region that corresponds to the {\it HST}\,/\,ACS footprint of the COSMOS field and represents the region of the S2COSMOS map that is closest to uniform with a median 1--$\sigma$ sensitivity of 1.2\,\mjpb\,over 1.6 sq.~degree. To extend our sensitivity to luminous, rare sources we define a {\sc supp} region that is contiguous to the {\sc main} survey and provides a further 1 sq.~degree of coverage at a median depth of 1.7\,\mjpb. Combined the {\sc main} and {\sc supp} regions correspond to a survey volume of 1.5\,$\times$\,10$^{8}$\,Mpc$^3$ at $z$\,=\,0.5--6.0 for sources brighter than $S_{850}$\,$\gsim$\,8\,mJy.

\item Above a signal--to--noise threshold of 4.0\,$\sigma$ and 4.3\,$\sigma$ we detect 1020 and 127 sources in the {\sc main} and {\sc supp} regions, respectively. Using jackknife maps of the S2COSMOS survey we estimate a false detection rate of 2\,$\pc$ for both samples, corresponding to 21 and 3 spurious detections integrated across the source catalogs. 

\item Simulations of the S2COSMOS survey are used to estimate the effect of flux boosting and the completeness level of our source catalogue. Using the results of these simulations we apply a correction for flux boosting on source-by-source basis based on the local noise and observed flux density for each source. The final S2COSMOS source catalogue contains sources with intrinsic flux densities of $S_{850}$\,=\,1.6--19.9\,mJy and we estimate that the {\sc main} and {\sc supp} survey regions are 50\,$\pc$ (90\,$\pc$) complete at $S_{850}$\,=\,4.4\,mJy (6.4\,mJy) and $S_{850}$\,=\,5.1\,mJy (9.1\,mJy), respectively. 

\item From the S2COSMOS source catalogue we construct the 850\,$\mu$m number counts from $S_{850}$\,=\,2--20\,mJy. The S2COSMOS differential number counts are well-modeled by a single Schechter function with best--fit parameters $N_{0}$\,=\,5000$^{+1300}_{-1400}$\,deg$^{-2}$, $S_{0}$\,=\,3.0$^{+0.6}_{-0.5}$\,mJy, and $\gamma$\,=\,1.6$^{+0.3}_{-0.4}$, and we find no evidence for an upturn in the bright--end of the counts due to strongly--lensed sources or local galaxies. The S2COSMOS differential counts are shown to be in agreement with the 850\,$\mu$m counts derived from S2CLS at the $<$\,1\,$\sigma$ significance level, and we conclude that cosmic variance does not strongly affect the 850\,$\mu$m population on scales of $\sim$\,0.5--3\,sq.~degree.

\item An average excess of 2.0\,$\pm$\,0.2 near--infrared--selected ($K_{s}$\,$\le$\,24.5 or [3.6\,$\mu$m]\,$\le$\,25.0\,mag) galaxies is measured within $<$\,$13''$ ($\sim$\,100\,kpc at $z$\,$\sim$\,2) of S2COSMOS sources, representing SMGs and\,/\,or associated galaxies. The ``excess'' arises from galaxies that lie at a median redshift of $z$\,=\,2.0\,$\pm$\,0.1 (significantly higher than a $K$--band--magnitude matched sample), and the distribution is in reasonable agreement with a sample of ALMA--identified, near--infrared--detected SMGs, confirming that the S2COSMOS survey has identified a population of high-redshift, starburst galaxies. 

\item We stack on the S2COSMOS map at the position of mass--selected galaxies, split into three subsets of stellar mass (log$_{10}$\,$M_{\star}$\,=\,10--12) and eight subsets in redshift ($z$\,=\,0--4). Stacked 850\,$\mu$m emission is detected from each subset in stellar mass and redshift at a SNR\,=\,4--30, corresponding to an average flux density of $S_{850}$\,=\,0.2--2.4\,mJy. We extend the stacking analysis to far--infrared wavelengths and identify emission at 100--500\,$\mu$m from each subset that this is in good agreement with that measured at 850\,$\mu$m. The far--infrared--to--sub--mm emission is modeled with a single--temperature mBB that transition to a power--law function in the mid--infrared to determine the far--infrared luminosities and the peak wavelength of the dust SED for each of our stacked

\item Using our SED fitting results we search for trends between the far--infrared luminosity and the peak wavelength of the dust SED for mass-selected galaxies at $z$\,=\,0--4. We identify a strong trend of decreasing peak wavelength (hotter characteristic dust temperature) with far--infrared luminosity (and redshift) that is broadly consistent with the properties of infrared--luminous sources at $z$\,=\,0--1, and a broad trend of decreasing peak wavelength with stellar mass, at a fixed far--infrared luminosity.

\item Finally, we construct a toy model built on empirical relations that broadly reproduces the observed trends in the L$_{\mathrm{FIR}}$--$\lambda_{\mathrm{peak}}$ relation for mass--selected sources, including a dependence on stellar mass. However, we caution that the empirical relations have significant associated uncertainties, are poorly calibrated at high redshift, and that the observed trends in the L$_{\mathrm{FIR}}$--$\lambda_{\mathrm{peak}}$ plane are sensitive to redshift evolution in the star-formation rate distribution of the galaxies in each of our stacked sub-samples. Despite these concerns, the observed stellar mass dependence in the L$_{\mathrm{FIR}}$--$\lambda_{\mathrm{peak}}$ plane, as identified in our stacks, appears robust and, under the assumptions of our toy model, provides circumstantial evidence for metallicity dependance in the dust--to--gas ratio to $z$\,$\sim$\,4.
\end{itemize}

\section*{Acknowledgements}
We sincerely thank the anonymous referee for providing a constructive report that improved the quality of this paper, as well as Matthieu Bethermin for providing theoretical predictions and Marco Viero for his guidance regarding the usage of {\sc simstack}. The James Clerk Maxwell Telescope is operated by the East Asian Observatory on behalf of The National Astronomical Observatory of Japan; Academia Sinica Institute of Astronomy and Astrophysics; the Korea Astronomy and Space Science Institute; the Operation, Maintenance and Upgrading Fund for Astronomical Telescopes and Facility Instruments, budgeted from the Ministry of Finance (MOF) of China and administrated by the Chinese Academy of Sciences (CAS), as well as the National Key R\&D Program of China (No. 2017YFA0402700). Additional funding support is provided by the Science and Technology Facilities Council of the United Kingdom and participating universities in the United Kingdom and Canada (ST/M007634/1, ST/M003019/1, ST/N005856/1). The James Clerk Maxwell Telescope has historically been operated by the Joint Astronomy Centre on behalf of the Science and Technology Facilities Council of the United Kingdom, the National Research Council of Canada and the Netherlands Organization for Scientific Research and data from observations undertaken during this period of operation is used in this manuscript. This research used the facilities of the Canadian Astronomy Data Centre operated by the National Research Council of Canada with the support of the Canadian Space Agency.  

JMS gratefully acknowledges support from the EACOA fellowship program. IRS acknowledges support from ERC Advanced Grant DUSTYGAL (321334), STFC (ST/P000541/1) and a Royal Science Wolfson Merit award. JLW acknowledges support from an STFC Ernest Rutherford Fellowship (ST/P004784/2). J.E.G. thanks the Royal Society.  Y.M acknowledges JSPS KAKENHI Grant Number 25287043, 17H04831, and 17KK0098. MJM acknowledges the support of the National Science Centre, Poland through the SONATA BIS grant 2018/30/E/ST9/00208 and POLONEZ grant 2015/19/P/ST9/04010, which received funding from the European Union's Horizon 2020 research and innovation programme under the Marie Sk{\l}odowska-Curie grant agreement No. 665778.

\bibliographystyle{apj} 
\bibliography{ref.bib}

\end{document}

%% file: tableshort.tex
\begin{table*}
 \centering
 \centerline{\sc Table 1: S2COSMOS Source Catalog}
\vspace{0.1cm}
 {%
 \begin{tabular}{lccccccc}
 \hline
 \noalign{\smallskip}
Name & Short ID & R.A. & Dec. & SNR & $S^{\mathrm{obs}}_{850}$\,$\pm$\,$\sigma_{\mathrm{inst}}$ & $S^{\mathrm{deb}}_{850}$\,$\pm$\,$\sigma_{\mathrm{total}}$$^{a}$ & Sample \\ 
 &  & (J2000) & (J2000) &  & (mJy) & (mJy) &  \\ 
\hline \\ [-1.9ex] 
S2COSMOS\,J100008+022611  & S2COS850.0001 & 10\,00\,08.05 & 02\,26\,11.6 & 28.4 & 16.8\,$\pm$\,0.6 & 16.8$^{+0.9}_{-1.0}$ & MAIN \\ 
S2COSMOS\,J100015+021549  & S2COS850.0002 & 10\,00\,15.52 & 02\,15\,49.6 & 22.3 & 13.5\,$\pm$\,0.6 & 13.3$^{+0.7}_{-1.4}$ & MAIN \\ 
S2COSMOS\,J100057+022013  & S2COS850.0003 & 10\,00\,57.16 & 02\,20\,13.6 & 19.5 & 13.0\,$\pm$\,0.7 & 12.8$^{+0.9}_{-1.3}$ & MAIN \\ 
S2COSMOS\,J100019+023203  & S2COS850.0004 & 10\,00\,19.79 & 02\,32\,03.6 & 19.1 & 13.2\,$\pm$\,0.7 & 13.2$^{+0.9}_{-1.1}$ & MAIN \\ 
S2COSMOS\,J100023+021751  & S2COS850.0005 & 10\,00\,23.93 & 02\,17\,51.6 & 19.0 & 10.5\,$\pm$\,0.6 & 10.3$^{+0.8}_{-1.0}$ & MAIN \\ 
S2COSMOS\,J095957+022729  & S2COS850.0006 & 09\,59\,57.37 & 02\,27\,29.6 & 18.2 & 12.1\,$\pm$\,0.7 & 12.0$^{+0.8}_{-1.5}$ & MAIN \\ 
S2COSMOS\,J100033+022559  & S2COS850.0007 & 10\,00\,33.40 & 02\,25\,59.6 & 16.0 & 9.4\,$\pm$\,0.6 & 9.2$^{+0.8}_{-1.1}$ & MAIN \\ 
S2COSMOS\,J100249+023255  & S2COS850.0008 & 10\,02\,49.26 & 02\,32\,55.1 & 15.4 & 20.4\,$\pm$\,1.3 & 19.6$^{+1.7}_{-1.5}$ & MAIN \\ 
S2COSMOS\,J100028+023203  & S2COS850.0009 & 10\,00\,28.73 & 02\,32\,03.6 & 14.6 & 10.1\,$\pm$\,0.7 & 9.9$^{+1.0}_{-1.3}$ & MAIN \\ 
S2COSMOS\,J100023+022155  & S2COS850.0010 & 10\,00\,23.53 & 02\,21\,55.6 & 14.4 & 7.9\,$\pm$\,0.5 & 7.7$^{+0.7}_{-1.1}$ & MAIN \\ 
...	& ... & ... & ... & ... & ... & ... &  ... \\
 \hline\hline \\  [0.5ex]  
 \end{tabular}
 \vspace{-0.5cm}
 \begin{flushleft}
 \footnotesize{ Example of the S2COSMOS source catalog, showing the 850--$\mu$m sources that are detected at the highest significance level, across the 2.6 sq.~degree S2COSMOS survey region. The full catalog is available in the online journal. $^{a}$ Deboosted flux density and associated uncertainty, including the contribution from instrumental and confusion noise.}
 \end{flushleft}
}
 \refstepcounter{table}
 \label{table:obs}
 \end{table*}

%% file: tablecounts.tex
\begin{table*}
 \centering
 \centerline{\sc Table 2: S2COSMOS Number Counts}
\vspace{0.1cm}
 {%
 \begin{tabular}{cccccc}
 \hline
 \noalign{\smallskip}
 $S_{850}$ & $N^{\textsc{m}}(>S_{850})$ & $N^{\textsc{m+s}}(>S_{850})$ & $S'_{850}$ & $dN^{\textsc{m}}/dS'_{850}$ & $dN^{\textsc{m+s}}/dS'_{850}$  \\ 
 (mJy) & (deg$^{-2}$) & (deg$^{-2}$) & (mJy) & (deg$^{-2}$\,mJy$^{-1}$) & (deg$^{-2}$\,mJy$^{-1}$)  \\ 
\hline \\ [-1.9ex] 
2.0 & 1920$^{+90}_{-90}$ & 1910$^{+90}_{-90.0}$    & 2.2 & 1370$^{+310}_{-280}$ & 1360$^{+300}_{-270}$ \\ 
2.3 & 1480$^{+70}_{-70}$ & 1470$^{+70}_{-70}$ 	   & 2.5 & 965$^{+184}_{-168}$ & 962$^{+179.3}_{-163}$ \\ 
2.7 & 1110$^{+50}_{-50}$ & 1110$^{+50}_{-50}$	   & 2.9 & 673$^{+111}_{-102}$ & 671$^{+107}_{-100}$ \\ 
3.1 & 822$^{+38}_{-38}$ & 813$^{+36}_{-35}$ 	   & 3.4 & 462$^{+67}_{-62}$ & 460$^{+64}_{-60}$ \\ 
3.6 & 588$^{+29}_{-28}$ & 579$^{+27}_{-26}$ 	   & 3.9 & 308$^{+41}_{-39}$ & 306$^{+40}_{-37}$ \\ 
4.2 & 406$^{+22}_{-22}$ & 398$^{+20}_{-19}$	 	   & 4.6 & 197$^{+25}_{-24}$ & 195$^{+24}_{-23}$ \\ 
4.9 & 271$^{+17}_{-16}$ & 264$^{+15}_{-14}$ 	   & 5.3 & 121$^{+16}_{-15}$ & 120$^{+15}_{-14}$ \\ 
5.7 & 175$^{+13}_{-12}$ & 169$^{+11}_{-11}$ 	   & 6.2 & 72.5$^{+10.7}_{-9.8}$ & 71.5$^{+9.7}_{-8.9}$ \\ 
6.6 & 108$^{+10}_{-10}$ & 103$^{+10}_{-10}$ 	   & 7.2 & 42.3$^{+7.3}_{-6.6}$ & 41.2$^{+6.3}_{-5.8}$ \\ 
7.7 & 61.9$^{+7.7}_{-7.2}$ & 58.7$^{+6.4}_{-6.0}$  & 8.3 & 23.2$^{+4.9}_{-4.5}$ & 22.2$^{+4.1}_{-3.8}$ \\ 
9.0 & 32.9$^{+5.7}_{-5.4}$ & 30.8$^{+4.6}_{-4.2}$  & 9.7 & 10.9$^{+3.1}_{-2.7}$ & 10.2$^{+2.6}_{-2.3}$ \\ 
10.4 & 17.1$^{+4.3}_{-3.7}$ & 15.9$^{+3.3}_{-2.8}$ & 11.2 & 4.3$^{+2.0}_{-1.6}$ & 4.1$^{+1.6}_{-1.2}$ \\ 
12.1 & 9.7$^{+3.3}_{-2.7}$ & 9.0$^{+2.4}_{-2.2}$   & 13.0 & 2.8$^{+1.4}_{-1.1}$ & 2.4$^{+1.1}_{-0.9}$ \\ 
14.1 & 3.7$^{+2.5}_{-1.6}$ & 4.3$^{+1.9}_{-1.5}$   & 15.2 & 1.1$^{+0.9}_{-0.7}$ & 1.0$^{+0.7}_{-0.5}$ \\ 
16.3 & 1.8$^{+1.9}_{-1.2}$ & 1.9$^{+1.4}_{-1.1}$   & 17.6 & 0.5$^{+0.6}_{-0.4}$ & 0.4$^{+0.5}_{-0.3}$ \\ 
19.0 & 0.6$^{+1.4}_{-0.8}$ & 0.4$^{+1.0}_{-0.3}$   & 20.5 & 0.2$^{+0.5}_{-0.3}$ & 0.1$^{+0.3}_{-0.1}$ \\ 

 \hline\hline \\  [0.1ex]  
 \end{tabular}
 \vspace{-0.2cm}
 \begin{flushleft}
 \footnotesize{Note: The cumulative and differential number counts at 850\,$\mu$m constructed from the S2COSMOS {\sc main} (M) and {\sc main+supp} (M+S) regions, corresponding to a survey area of 1.6 and 2.6\,sq.~degree, respectively. Differential S2COSMOS counts are constructed in flux bins centered at an intrinsic 850\,$\mu$m flux, $S'_{850}$, with the cumulative counts measured at an intrinsic flux $>$\,$S_{850}$. There is excellent agreement between the counts constructed from each of our survey regions and, as such, throughout this work we adopt the counts measured from the combined {\sc main+supp} survey.}
 \end{flushleft}
 }
 \refstepcounter{table}
 \label{table:obs}
 \end{table*}